\documentclass[12pt,tightenlines]{revtex4} 
\usepackage{amssymb,amsmath}
\usepackage{psfrag}
\usepackage[dvips]{graphicx}
\usepackage{verbatim}
\usepackage{color}
\definecolor{darkblue}{rgb}{0,0,0.5}
\usepackage[dvips, unicode, colorlinks, bookmarks=true, citecolor=darkblue, linkcolor=darkblue, urlcolor=darkblue]{hyperref}

\newcommand{\hc}{\mathrm{h.c.}}
\begin{document}

\title{Quantum variational measurement and the ``optical lever'' intracavity topology of gravitational-wave detectors}

\author{F.Ya.Khalili}

\email{farid@hbar.phys.msu.ru}
\affiliation{Physics Faculty, Moscow State University, Moscow 119992, Russia}


\begin{abstract}
  The intracavity topologies of laser gravitational-wave detectors are the promising way to obtain sensitivity of these devices significantly better than the Standard Quantum Limit (SQL). The most challenging element of the intracavity topologies is the \emph{local meter} which has to monitor position of a small ($1\div10$ gram) local mirror and which precision defines the sensitivity of the detector. 

  To overcome the SQL, the quantum variational measurement can be used in the local meter. In this method a frequency-dependent correlation between the meter back-action noise and measurement noise is introduced, which allows to eliminate the back-action noise component from the meter output signal. This correlation is created by means of an additional filter cavity.

  In this article the sensitivity limitations of this scheme imposed by the optical losses both in the local meter itself and in the filter cavity are estimated. It is shown that the main sensitivity limitation stems from the filter cavity losses. In order to overcome it, it is necessary to increase the filter cavity length. In a preliminary prototype experiment about 10 meter long filter cavity can be used to obtain sensitivity approximately $2\div3$ times better than the SQL. For future QND gravitational-wave detectors with sensitivity about ten times better than the SQL, the  filter cavity length should be within kilometer range.
\end{abstract}

\maketitle




\section{Introduction}

It has been known for almost 40 years that position of a quantum object can not be monitored continuously with arbitrarily high precision. This limitation was called the Standard Quantum Limit (SQL) \cite{67a1eBr}. It arises because in position measurements two different kinds of noise sources exist: the measurement noise which is added by the meter to its input signal, and the back-action noise which perturbs the monitored object momentum and thus alters in a random way the values of its position in future moments of time. Due to the Heisenberg uncertainty relation these noises can not be canceled simultaneously.

If the goal is not the position measurement itself but the detection of an external force acting on a test object, then this limitation can be evaded by using more sophisticated measurement procedures. In particular, several authors \cite{Unruh1982, 87a1eKh, JaekelReynaud1990, Pace1993, 96a2eVyMa} independently proposed the principle of detection of a classical force acting on a test mass, which uses frequency-dependent correlation between the measurement noise and the back-action noise and thus allows to eliminate the back-action noise from the meter output. Hereinafter this method, following paper \cite{96a2eVyMa}, will be referred to as a {\em variational measurement}.

In the article \cite{02a1KiLeMaThVy} a possible implementation of variational measurement in optical interferometric position meters and, in particular, in laser gravitational-wave detectors was considered. The correlation between the measurement noise and back-action noise can be introduced in this case by using homodyne detector which measures weighted sum of the phase and amplitude quadrature components of the output optical beam with weight factors depending on the local oscillator phase. However, this method on its own allows to create frequency-independent correlation only, which allows to compensate the back action at one given frequency.

In order to create the frequency-dependent correlation, it was proposed in the article \cite{02a1KiLeMaThVy} to reflect the interferometer output beam sequentially from two Fabri-Perot cavities with suitably adjusted bandwidths and detunings, thus introducing a frequency-dependent phase shift in the output beam.

The filter cavities bandwidths have to be of the same order of magnitude as the signal frequency. In case of terrestrial laser gravitational wave detectors the characteristic scale of this frequency is $\Omega\sim 10^2\div 10^3\,{\rm s}^{-1}$\,. These frequencies are very low compared to the optical ones, therefore long filter cavities with very high-reflectivity mirrors should be used. In the article \cite{02a1KiLeMaThVy} the filter cavities with the same length as the main interferometer cavities (4\,Km), placed  side-by-side with the latter ones in the same vacuum chamber, were considered.

This design has, however, some disadvantages. The most important one is inherent in all large-scale laser interferometric position meters: very high optical power which has to circulate in the interferometer arms. For example, in planned Advanced LIGO gravitational-wave detector, almost one megawatt is necessary just in order to reach the SQL \cite{WhitePaper1999}. The second disadvantage is specific for the scheme proposed in \cite{02a1KiLeMaThVy}: photons scattering from the main cavities into the filter ones will distort the back-action compensation mechanism \cite{Whitecomb_private}. It should be noted that in contrast with huge optical power in the main arms, the optical field quantum state in the filter cavities should be close to vacuum one.

\begin{figure*}
  \psfrag{T1}{}\psfrag{T2}{}\psfrag{T3}{}\psfrag{T4}{}
  \psfrag{T5}{}\psfrag{T6}{}\psfrag{T7}{}\psfrag{TL}{}
  \psfrag{A}{}\psfrag{B}{}\psfrag{A1}{}\psfrag{B1}{}
  \psfrag{C}[lb][lb]{{\sf C}}\psfrag{D}[rt][rt]{{\sf D}}
  \includegraphics[width=0.45\textwidth]{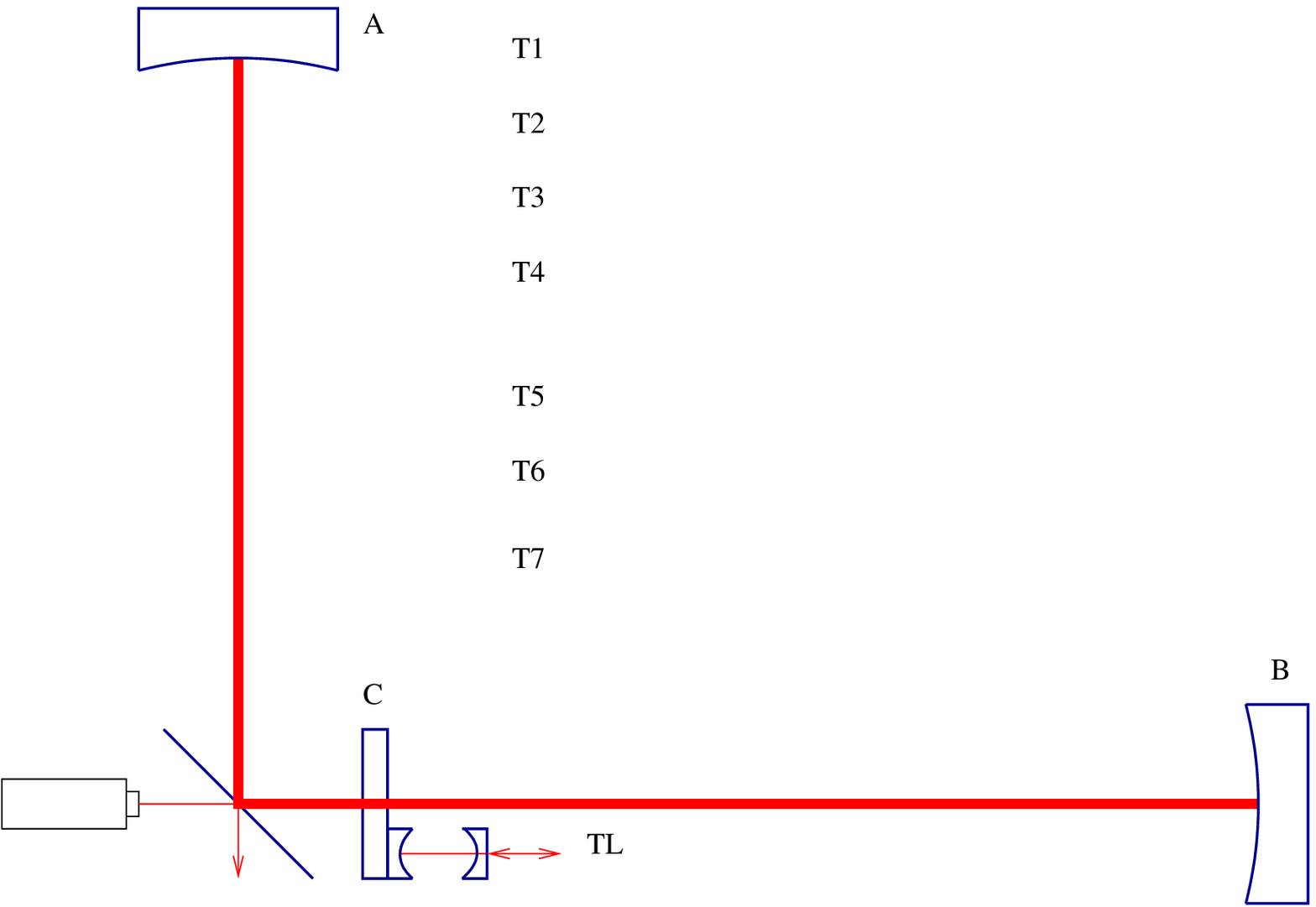}\hspace{\fill}
  \includegraphics[width=0.45\textwidth]{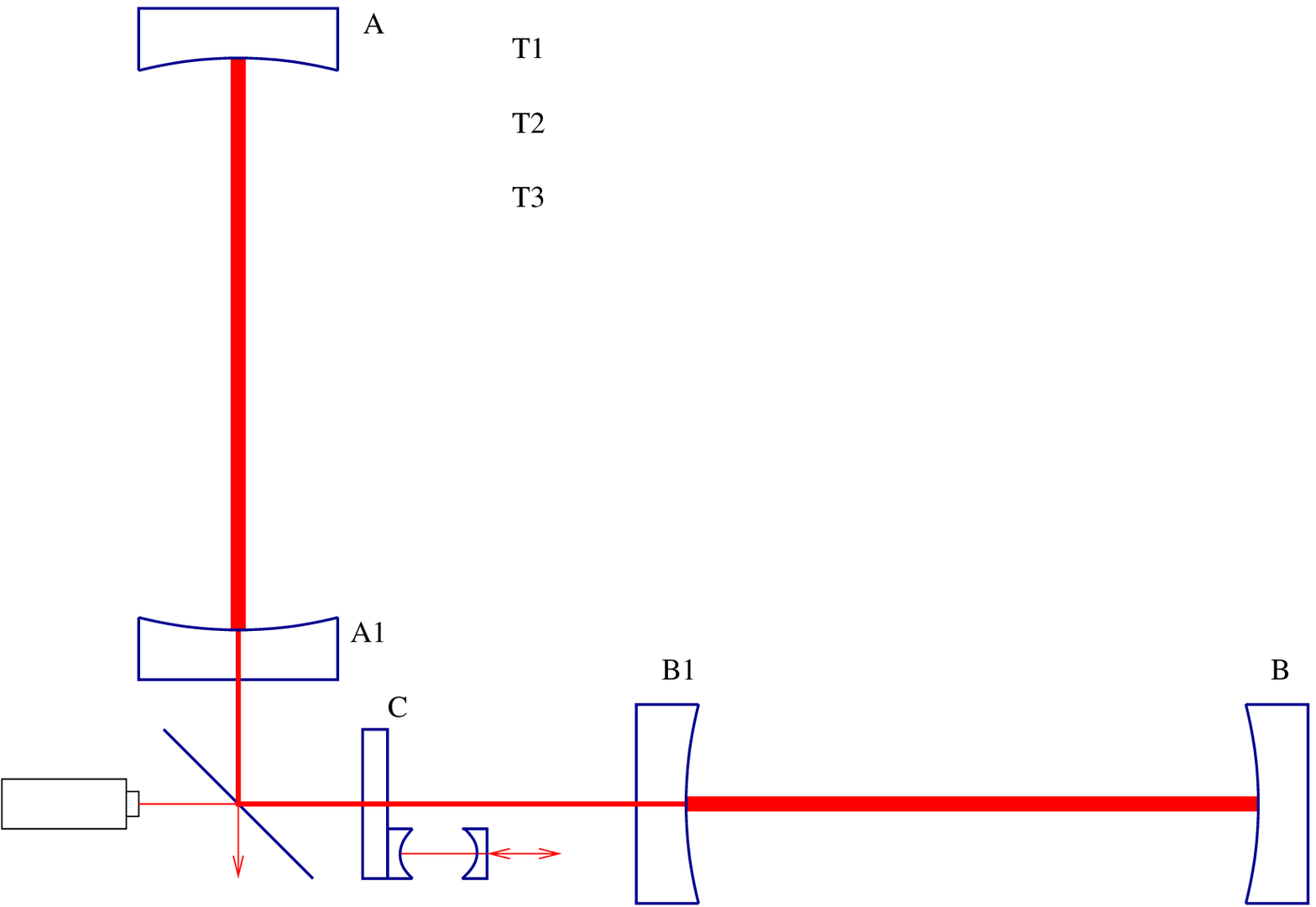}
  \caption{The ``optical bars'' (left) and ``optical lever'' (right)
  intracavity schemes.}\label{fig:optbar}
\end{figure*}

Alternative design of large-scale optical gravitational-wave detectors based on the Quantum Non-Demolition intracavity readout, the {\em intracavity} scheme, was proposed in the article \cite{96a2BrKh} and the possible implementation of this scheme (the {\em optical bars} topology) was proposed in the article \cite{97a1BrGoKh}. The basic idea of this topology is the following [see Fig.\,\ref{fig:optbar}(left)]: instead of direct measurement of the end mirrors displacement, transfer it to the displacement of an additional local mirror {\sf C} by means of optical rigidity and then measure this  displacement using an additional small-scale {\em local meter}. 

The main advantage of this topology is the significant reduction of requirements for circulating optical power value (see papers \cite{03a1Kh, 06a1DaKh}). Another advantage is that it is possible in this scheme , using two additional mirrors ${\sf I_1, I_2}$, to amplify the signal displacement by the factor of $\sqrt{M/m}$, where $M, m$ are the end and the local mirrors masses, correspondingly \cite{02a1Kh} [the {\em optical lever} topology, see Fig.\,\ref{fig:optbar}(right)]. However, the SQL is also inversely proportional to the square root of test mass. Therefore this amplification allows to reduce influence of miscellaneous noises of technical origin, but does not allow to obtain sensitivity better than the SQL by itself.

One of the most promising methods to overcome the SQL in this topology is to use the variational measurement in the local meter \cite{06a1DaKh}. This scheme, while based on the same principle as the scheme of the article \cite{02a1KiLeMaThVy}, has one important difference. The local meter cavity (or cavities) should be relatively short (with length $l\lesssim1\mathrm{m}$). It is known (see, for example, consideration in \cite{06a1KhLaVy}) that an optical interferometric position meter can provide sensitivity better than the SQL only if this meter effective loss factor ${\cal A}$ (see the main notations list in Table\,\ref{tab:notations}) is small. Simple estimates show that in this case even for the best mirrors, available now, the meter cavity bandwidth will be much larger than characteristic gravitational-wave signal frequencies:
\begin{equation}\label{wb_cond} 
  \gamma=\frac{c(T^2+A^2)}{4l} \gg \Omega \sim 10^2\div 10^3\,{\rm s}^{-1} \,,
\end{equation} 
It can be shown (see below) that in this case only one (instead of two) filter cavity is required.

In this article the described above variational measurement scheme with single filter cavity is analyzed. In Sec.\,\ref{sec:criterion} a new, more consistent and ``fair'' (in author's opinion) criterion of the SQL overcoming in wide band is proposed. In Sec.\,\ref{sec:sensitivity} the possible meter topologies are discussed and the sensitivity limitations imposed by optical losses both in the meter and in the filter cavities, are calculated. 

Other noise sources, most notably the thermal noises in the mirrors coating, bulk, and suspension, are not considered in detail in this paper because these noises influence depends significantly on specific values of the experimental setup parameters. The suspension noise was estimated in the article \cite{01a1BrKhVo}, where it has been shown not to prevent from obtaining the sensitivity about ten times better than the SQL.  The mirrors coating and bulk thermal noises are discussed in Sec.\,\ref{sec:mirrors} in brief.

\begin{table*}[t]
  \begin{tabular}{|c|c|l|}
    \hline
      Quantity    & Value for estimates                 & Description \\
    \hline 
      $c$         & $3\times10^8\,\mathrm{m/s}$         & Speed of light \\
      $m$         &                                     & Test mass \\
      $\Omega_c$  &                                     & The cut-off frequency \\
      $\omega_p$  & $1.8\times10^{15}\,\mathrm{s}^{-1}$ & Optical pumping frequency \\
      $T^2$       &                                     & Meter cavity input mirror transmittance \\
      $A^2$       & $10^{-5}$                           & Meter cavity losses per bounce \\
      ${\cal A} = A^2/T^2$ &                               & Effective loss factor of the meter cavity \\
      $W$         &                                     & Optical power circulating in the meter cavity\\
      $l_f$       &                                     & Filter cavity length \\
      $T_f^2$     &                                     & Filter cavity input mirror transmittance \\
      $A_f^2$     & $10^{-5}$                           & Filter cavity losses per bounce \\
    \hline  
  \end{tabular}
  \caption{Main notations used in this paper.}\label{tab:notations}
\end{table*}

\section{Criterion of the Standard Quantum Limit overcoming}\label{sec:criterion}

\subsection{Discussion of the criterion}

The sum noise spectral density of the ``ordinary'' (SQL-limited) position meter is equal to \cite{92BookBrKh, 03a1BrGoKhMaThVy}
\begin{equation}\label{S_SQL_1} S(\Omega) = S_x + \frac{S_F}{m^2\Omega^4} \,. \end{equation} 
Here $S_x$ and $S_F$ are the measurement noise and back-action noise spectral densities, correspondingly, which satisfy the following uncertainty relation:
\begin{equation}\label{SxSF} S_xS_F = \frac{\hbar^2}{4} \,. \end{equation} 
(in general, it has an inequality form but we suppose here that the meter is quantum-noise limited one so its noises are as small as possible). 

In case of optical interferometric position meters measurement noise (also known in this case as shot noise) is inversely proportional to optical power, and back action noise (also known as radiation-pressure noise) is directly proportional to it. If resonant optical pumping is used, condition (\ref{wb_cond}) is fulfilled and the phase quadrature amplitude of the output beam is measured, then the sum noise spectral density has the simplest form (\ref{S_SQL_1}); for the explicit values of $S_x$ and $S_F$, see Eqs.\,(\ref{Sx_SF_0}).

Using the uncertainty relation (\ref{SxSF}), Eq.\,(\ref{S_SQL_1}) can be rewritten as follows:
\begin{equation}\label{S_SQL_2} 
  S(\Omega) = \frac{\hbar}{2m\Omega_c^2}\left(1 + \frac{\Omega_c^4}{\Omega^4}\right) \,,
\end{equation} 
where
\begin{equation}\label{Omega_c_SQL} 
  \Omega_c = \sqrt{\frac{\hbar}{2mS_x}} = \sqrt{\frac{32\omega_pW}{mc^2T^2}} \,.
\end{equation} 
is the meter {\em cut-off frequency}, {\em i.e.} the characteristic frequency where, due to back-action noise, the sum noise spectral density increases by factor of 2 compared to the asymptotic value at $\Omega\to\infty$:
\begin{equation}\label{Omega_c} S(\Omega_c) = 2S(\Omega)|_{\Omega\to\infty} \,.\end{equation} 

\begin{figure}
  \psfrag{Omega}[cc][cc]{$\Omega$}
  \psfrag{S}[cc][cc]{$mS(\Omega)/\hbar$}
  \psfrag{SQL}{\rotatebox{-45}{SQL}}
  \includegraphics[width=0.45\textwidth]{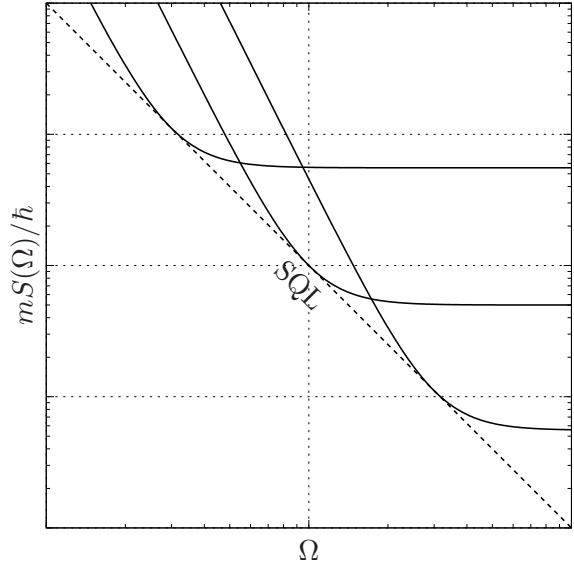}
  \caption{The noise curves family of SQL-limited position meters; the common envelope of these curves form the Standard Quantum Limit.}\label{fig:SQL}
\end{figure}

Eq.\,(\ref{S_SQL_2}) describes a family of functions parametrized by the frequency $\Omega_c$ (see Fig.\,\ref{fig:SQL}). Common envelope of this family
\begin{equation}\label{S_SQL_env} S_{\rm SQL}(\Omega) = \frac{\hbar}{m\Omega^2} \end{equation} 
is usually considered as the sub-SQL sensitivity borderline. Typically, sensitivity of schemes which allow to overcome the SQL is compared with this envelope. 

\begin{figure}
  \psfrag{Omega}[cc][cc]{$\Omega$}
  \psfrag{Omega1}[cc][lc]{$\Omega_1$}
  \psfrag{Omega2}[cc][lc]{$\Omega_2$}
  \psfrag{S}[cc][cc]{$mS(\Omega)/\hbar$}
  \psfrag{1}[lb][cc]{(a)}
  \psfrag{2}[lb][cc]{(b)}
  \psfrag{3}[lb][cc]{(c)}
  \psfrag{SQL}{\rotatebox{-45}{SQL}}
  \includegraphics[width=0.45\textwidth]{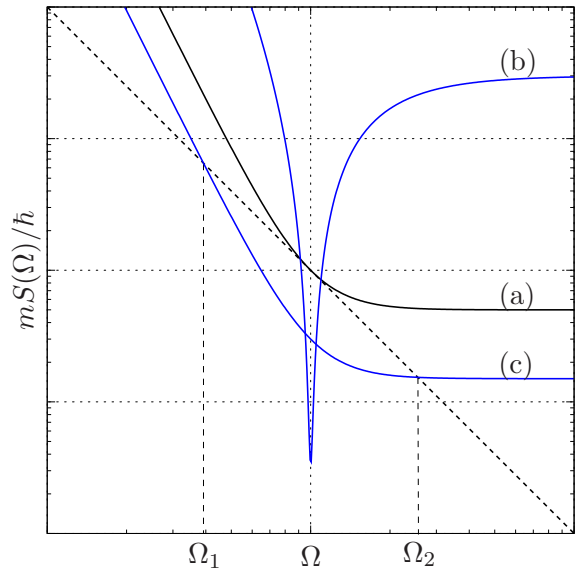}
  \caption{Noise curves of: (a) --- SQL-limited meter; (b) --- meter which overcomes the SQL in narrow band; (c) --- meter which overcomes the SQL in wide band. Curves (a) and (c) have the same value of the cut-off frequency $\Omega_c$.}\label{fig:sub_SQL}
\end{figure}

This comparison suffers from some ``unfairness'' because an {\em envelope} is compared with noise curve of a {\em specific meter}. It can be used in narrow-band cases as shown in Fig.\,\ref{fig:sub_SQL}, curve (b), because within a narrow frequency band the envelope (\ref{S_SQL_env}) can always be approximated by one of the curves from the family (\ref{S_SQL_2}) [see curve (a) in Fig.\,\ref{fig:sub_SQL}]. The noise curve (b) can be obtained, for example, using the frequency-independent noises correlation.

On the other hand, consider a hypothetic meter with the sum noise spectral density described by curve (c) which is obtained from curve (a) by multiplying by some factor $\xi^2<1$ (see again Fig.\,\ref{fig:sub_SQL}). From the point of view of the above noted criterion, the SQL is overcome here within the limited frequency band $\Omega_1<\Omega<\Omega_2$ only. On the other hand, curve (c) corresponds to $\xi^{-2}$ times  smaller noise spectral density than that of the SQL-limited meter [curve (a)] for all frequencies.

In the current paper, we will characterize the sensitivity by the factor
\begin{equation}\label{my_crit}
  \xi^2 = \left.\frac{2m\Omega_c^2S(\Omega)}{\hbar}\right|_{\Omega\to\infty} \,,
\end{equation} 
where the cut-off frequency $\Omega_c$ is defined by Eq.\,(\ref{Omega_c}). It is easy to see that $\xi^2=1$ for all SQL-limited position meters with sum noise spectral densities described by Eq.\,(\ref{S_SQL_1}).

\subsection{Frequency-independent correlation}

Consider simple example: position meter with the sum noise spectral density equal to
\begin{equation}\label{S_SQL_corr} 
  S(\Omega) = S_x - \frac{2S_{xF}}{m\Omega^2}+ \frac{S_F}{m^2\Omega^4} \,,
\end{equation}
where spectral densities $S_x$, $S_{xF}$, and $S_F$ satisfy the following uncertainty relation:
\begin{equation}\label{SxSF_corr} S_xS_F - S_{xF}^2 = \frac{\hbar^2}{4} \,. \end{equation} 
For example, the optical interferometric position meter considered above, will have the sum noise spectral density of the form (\ref{S_SQL_corr}) if not phase but some arbitrary quadrature amplitude of the output beam is measured and thus frequency-independent correlation of the measurement noise and back-action noise is introduced [see Eqs.\,(\ref{S_sum}, \ref{Sx_SF_res})].

The cut-off frequency defined by the condition (\ref{Omega_c}) in this case is equal to:
\begin{equation}
  \Omega_c = \sqrt{\frac{\sqrt{\hbar^2/4 + 2S_{xF}^2} - S_{xF}}{mS_x}} \,.
\end{equation} 
Therefore,
\begin{equation}
  \xi^2 = \frac{2}{\hbar}\left(\sqrt{\hbar^2/4 + 2S_{xF}^2} - S_{xF}\right) \,.
\end{equation} 
The minimum of this value corresponds to
\begin{equation}\label{S_xF_opt} S_{xF} = \frac{\hbar}{2\sqrt{2}}\,, \end{equation} 
and is equal to
\begin{equation}\label{xi_corr} \xi^2 = \frac{1}{\sqrt{2}} \,, \end{equation}
The corresponding cut-off frequency is equal to
\begin{equation}\label{Omega_c_corr}
  \Omega_c = \sqrt{\frac{\hbar}{2\sqrt{2}mS_x}} = \sqrt{\frac{32\sqrt{2}\omega_pW}{3mc^2T^2}} \,.
\end{equation} 

\begin{figure}
  \psfrag{Omega}[cc][cc]{$\Omega$}
  \psfrag{S}[cc][cc]{$mS(\Omega)/\hbar$}
  \includegraphics[width=0.45\textwidth]{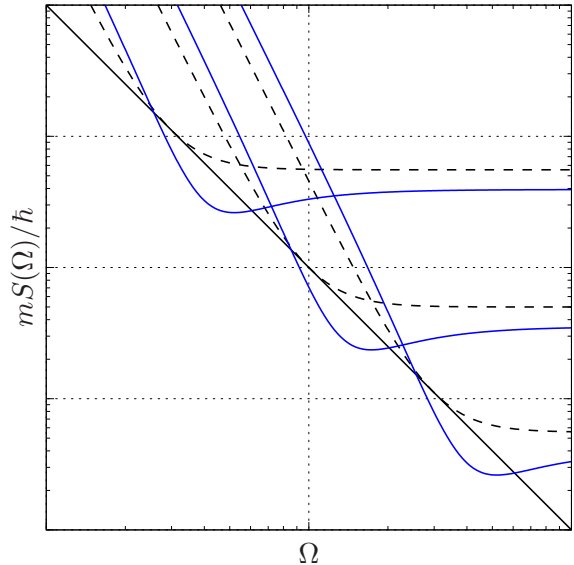}
  \caption{Solid lines: the noise curves family of position meters with optimized [see Eq.\,(\ref{S_xF_opt})] frequency-independent noises correlation; dashed lines: the noise curves for corresponding SQL-limited meters.}\label{fig:const_phi}
\end{figure}

It follows from this simple consideration that frequency-independent noises correlation allows to decrease the sum noise spectral density by the factor of $\sqrt{2}$ in wide band. This result is illustrated by Fig.\,\ref{fig:const_phi} where spectral densities (\ref{S_SQL_corr}) optimized using condition (\ref{S_xF_opt}), are plotted for different values of $\Omega_c$.

It should be also noted that there is another method of obtaining the sum noise spectral density of the form (\ref{S_SQL_corr}): use an ordinary position meter (with non-correlated noises), attached to a harmonic oscillator (instead of a free mass). Combination of these two methods ({\it i.e.} meter with cross-correlated noises attached to a harmonic oscillator) is also possible [see Eqs.\,(\ref{S_sum}, \ref{Sx_SF_eff})].
 
In quantum experiments the pondermotive rigidity \cite{67a1eBrMa, 70a1eBrMa} should be used in order to turn test mass into harmonic oscillator because it has extremely low level of internal mechanical noise \cite{99a1BrKh, 01a1BrKhVo}. In a single Fabry-Perot cavity the pondermotive rigidity can be created by using detuned (non-resonant) pumping, and in the signal-recycled Michelson---Fabry-Perot topology, similar to one proposed for the Advanced LIGO, one can obtain it by adjusting the signal recycling mirror position \cite{Buonanno2001, Buonanno2002, 05a1LaVy}.

\section{The variational measurement}\label{sec:sensitivity}

\subsection{Discussion of the meter topology}\label{sec:scheme}

\begin{figure}
  \psfrag{C}[rb][lb]{}
  \psfrag{m}[lb][lb]{$m$}
  \psfrag{W}[cb][lb]{$W$}
  \psfrag{D}[lc][lb]{\parbox{0.15\textwidth}{\sf\small To homodyne\\detector}}
  \psfrag{F}[ct][lb]{\sf Filter cavity}
  \psfrag{M}[ct][lb]{\sf Meter cavity}
  \includegraphics[width=0.48\textwidth]{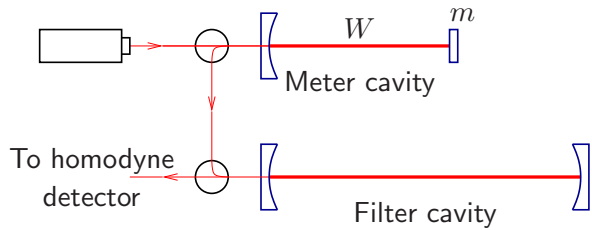}
  \caption{Conceptual model of variational measurement.}\label{fig:cscheme}
\end{figure}

In the simplest case variational measurement based meter can consist of two Fabry-Perot cavities: the meter cavity, with one of the mirrors attached to the test mass, and the filter one (see Fig.\,\ref{fig:cscheme}). The pumping laser beam is reflected first from the meter cavity. The output beam, which now carries in its phase information on the test mass position, is reflected from the filter cavity and, finally, is detected by the standard homodyne detector scheme. For both cavities, in order to separate the  output beams from the input one, some optical circulator schemes (the circles in Fig.\,\ref{fig:cscheme}) have to  be used (another option is to use ring cavities; however, in this case three mirrors instead of two have to be used in each of the cavities, which increases optical losses).

This scheme, while useful for understanding the general idea, can not be used in practice because in this case almost all pumping power (excluding the small fraction absorbed in the meter cavity) passes through the filter cavity. In this case all the uncertainties of filter cavity mirrors surfaces positions, caused by radiation  pressure fluctuations, Brownian noises, {\em etc}, will create additional fluctuational phase shifts in the output beam, masking the signal phase shift and thus degrading the sensitivity.

\begin{figure*}
  \psfrag{C}[cb][lb]{\sf C}
  \psfrag{D}[lc][lb]{\parbox{0.15\textwidth}{\sf\small To homodyne\\detector}}
  \psfrag{m}[cb][lb]{$m$}
  \psfrag{W}[cb][lb]{$W/2$}
  \psfrag{M}[ct][lt]{\sf Meter cavities}
  \psfrag{F}[ct][lb]{\sf Filter cavity}
  \psfrag{P}[lb][lb]{\sf PRC}
  \hspace{\fill}
  \includegraphics[width=0.48\textwidth]{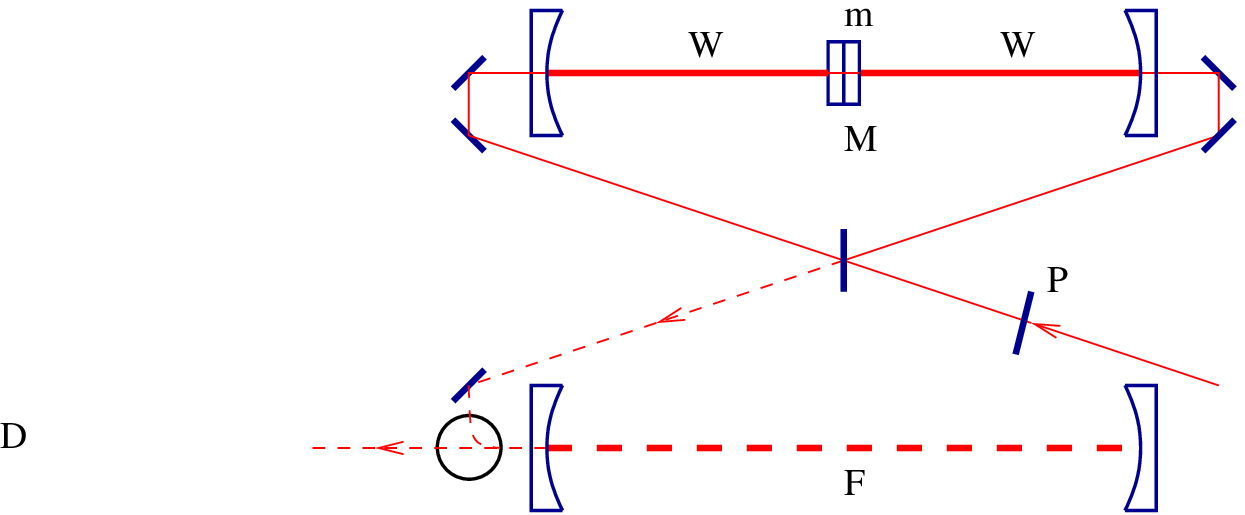}\hspace{\fill}
  \includegraphics[width=0.3\textwidth]{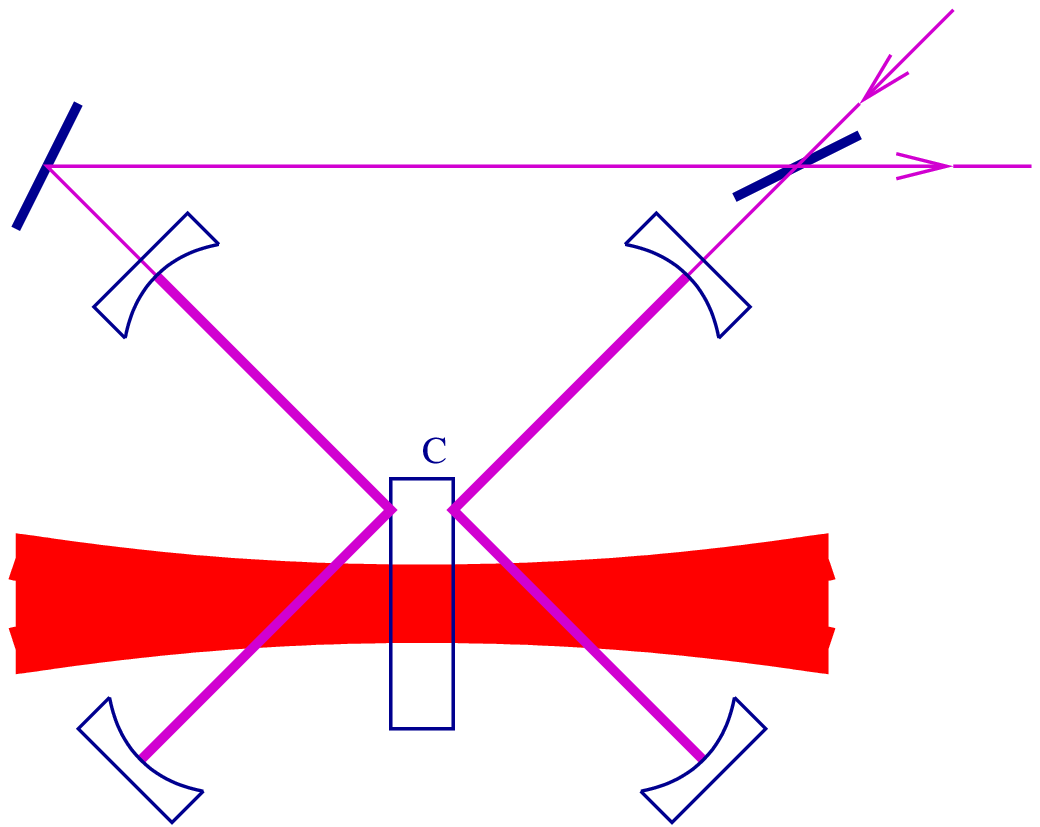}
  \hspace{\fill}
  \caption{Left: the possible topology of small-scale variational measurement setup ({\sf PRC} --- power recycling mirror which may be necessary to compensate the detuning created by test mass $m$ transmittance); right: the possible layout of the local meter cavities in optical bars/optical lever intracavity topology of gravitational-wave detectors.}\label{fig:rscheme}
\end{figure*}

Therefore, the balanced scheme of the meter interferometer with dark port output (similar to the standard Michelson---Fabry-Perot topology of laser gravitational wave detectors) should be used. The variant of such a scheme which can be used in small-scale prototype experiments (it has been first proposed for the pondermotive squeezing experiment \cite{Corbitt2004, 06a1CoChKhOtVyWhMa}) is shown in Fig.\,\ref{fig:rscheme}(left). Here two end mirrors of the arm Fabry-Perot cavities are glued together forming one test mass. The possible layout of the meter cavities suitable for the local meter of the optical bars/optical lever intracavity topology \cite{06a1DaKh} is shown in Fig.\,\ref{fig:rscheme}(right). Here {\sf C} is the local mirror (see Fig.\,\ref{fig:optbar}).

The scheme shown in Fig.\,\ref{fig:rscheme}(left) has an additional advantage not mentioned in papers \cite{Corbitt2004, 06a1CoChKhOtVyWhMa}. It allows to use end mirrors with moderate value of the reflection factor, comparable to input mirrors one. An additional short Fabry-Perot cavity created by the mirrors front surfaces have to be tuned in anti-resonance (compare with the double-mirror idea proposed in \cite{05a1Kh}).

Due to topology, the arm cavities finesse does not degrade in this case. Instead, a small optical coupling between the arm cavities arises, which does not affect the scheme functioning (see more detailed analysis in App.\,\ref{app:corbitt_scheme}). At the same time, moderate reflectivity mirrors (with smaller number of dielectric coating layers) have much smaller level of the coating thickness Brownian and thermoelastic fluctuations which are among the main factors which limit the gravitational-wave detectors sensitivity. 

It is known (see, for example, detailed consideration in \cite{Buonanno2003}) that analysis of the balanced Michelson---Fabry-Perot topology can be reduced to the analysis of the simpler scheme (Fig.\,\ref{fig:cscheme}). It is shown in App.\,\ref{app:corbitt_scheme} that this is also valid for the topology presented in Fig.\,\ref{fig:rscheme} with partly transparent end mirrors. The only significant difference here is that end mirrors transparency creates a relatively small detuning between the symmetric and anti-symmetric optical modes of the scheme, which can be compensated, for example, by an additional power recycling mirror {\sf PRC}. Therefore, the scheme  of Fig.\,\ref{fig:cscheme} will be considered below in this paper (presuming that the dark port condition is provided at the filter cavity input).

\subsection{The sensitivity}

The sensitivity of variational measurement scheme discussed above is calculated in Appendix \ref{app:var_corr}, see Eqs.\,(\ref{KLMTV_g_xi_res}, \ref{KLMTVs_optD}, \ref{KLMTV_g_xi_det}). In case of resonance tuned meter cavity, the factor $\xi$ introduced in Sec.\,\ref{sec:criterion} is equal to:
\begin{equation}\label{KLMTVs_xi_res} 
  \xi \approx \left(q_f + \sqrt{q_f^2 + q/2}\right)^{1/3}\,. 
\end{equation} 
The small optimal detuning (\ref{KLMTVs_optD}) allows to improve the sensitivity, giving
\begin{equation}\label{KLMTVs_xi_det}
  \xi \approx \left[\frac{1}{2}\left(q_f + \sqrt{q_f^2 + q}\right)\right]^{1/3} \,.
\end{equation}
Here
\begin{align}\label{KLMTVs_ggf}
  q &= \dfrac{mc^2\Omega_c^2A^2}{16\omega_pW} \,, &
  q_f &= \dfrac{cA_f^2}{2\sqrt{2}l_f\Omega_c} 
\end{align}
are the effective loss factors of the meter and filter cavities.

While losses in both the meter cavity $A^2$ and filter cavity $A_f^2$ appear in these equations, they limit the sensitivity in a different way. The influence of the losses in the meter cavity 
can be reduced by the increase of the cavity input mirror transmittance. The price for this is the increase of the optical power $W$ in the cavity (see the factor $q$). Estimates show that at contemporary technological level the meter cavity losses influence can be suppressed significantly. Really, if $A^2\approx10^{-5}$, then
\begin{equation}
  q \approx 1.2\times10^{-5}\times\left(\frac{1\mathrm{kW/g}}{W/m}\right)
    \times\left(\frac{\Omega_c}{2\pi\times100\,\mathrm{s}^{-1}}\right)^2
\end{equation} 
Using a several kilowatts of the circulating power and a small test mass $m\sim\,1\mathrm{gr}$ (these values are close to ones planned for the pondermotive squeezing experiment \cite{Corbitt2004, 06a1CoChKhOtVyWhMa}), it is possible to obtain $q\lesssim10^{-6}$. This value of $q$ allow, in principle, to reach the sensitivity sensitivity ten times better than the SQL ($\xi\lesssim0.1$).

On the other hand, the filter cavity bandwidth is fixed by variational measurement conditions [see Eqs.(\ref{KLMTVs_parms0(b)}, \ref{KLMTVs_parms0(c)}, \ref{KLMTVs_parms(b)})]. Therefore, the filter cavity losses influence can be reduced for given values of $A_f$ and $\Omega_c$ only by increase of the filter cavity length $l_f$, see the factor $q_f$. If $A_f^2\approx10^{-5}$, then 
\begin{equation}
  q_f \approx 1.7\times\left(\frac{1\mathrm{m}}{l_f}\right)
    \times\left(\frac{2\pi\times100\,\mathrm{s}^{-1}}{\Omega_c}\right) \,.
\end{equation} 
Therefore, for small-scale ($l_f\lesssim 10\,\mathrm{m}$) filter cavities only modest sensitivity gain ($\xi\gtrsim0.3$) can be obtained. However, it can be considered as sufficient for preliminary demonstration experiments. In the future QND gravitational-wave detectors, kilometer-scale filter cavities have to be used in order to obtain sensitivity significantly better than the SQL ($\xi\lesssim0.1$). 

It have to be noted that in intracavity readout topologies the problem of photon scattering into the filter cavity mentioned in the Introduction, can be avoided by using different optical frequencies in main interferometer and in the local meter \cite{Whitecomb_private}.

\begin{figure}
  \psfrag{xi}[ct][cb]{$\xi$}
  \psfrag{j}[ct][cc]{$W/m\,\mathrm{[W/g]}$} 
  \psfrag{x}[cc][lt]{$\otimes$} 
  \psfrag{a}[cb][cb]{(a)}
  \psfrag{b}[ct][cb]{(b)}
  \psfrag{c}[cb][cb]{(c)}
  \psfrag{d}[rt][lb]{(d)}
  \includegraphics[width=0.45\textwidth]{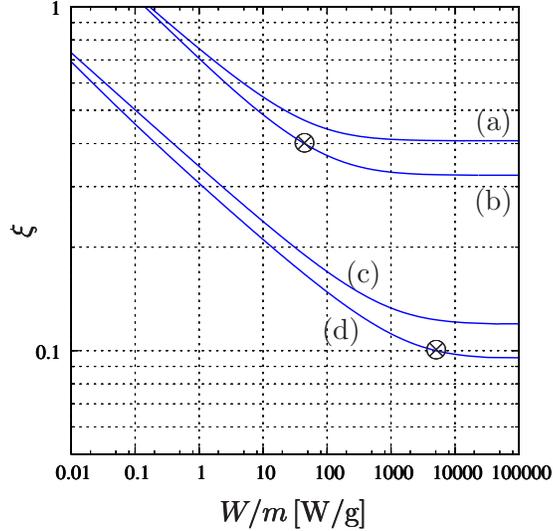}\bigskip
  \caption{Sensitivity dependence on the optical power: (a) resonant pumping, $l_f=10\,\mathrm{m}$; (b) optimal detuning, $l_f=10\,\mathrm{m}$; (c) resonant pumping, $l_f=4\,\mathrm{km}$; (d) optimal detuning, $l_f=4\,\mathrm{km}$.}\label{fig:varmeas}
\end{figure}  

\begin{figure}
  \psfrag{xi}[ct][cb]{$\sqrt{2m\Omega_c^2S(\Omega)/\hbar}$}
  \psfrag{Omega}[cc][cc]{$\Omega/\Omega_c$} 
  \psfrag{a}[cb][cb]{(a)}
  \psfrag{b}[cb][cb]{(b)}
  \includegraphics[width=0.45\textwidth]{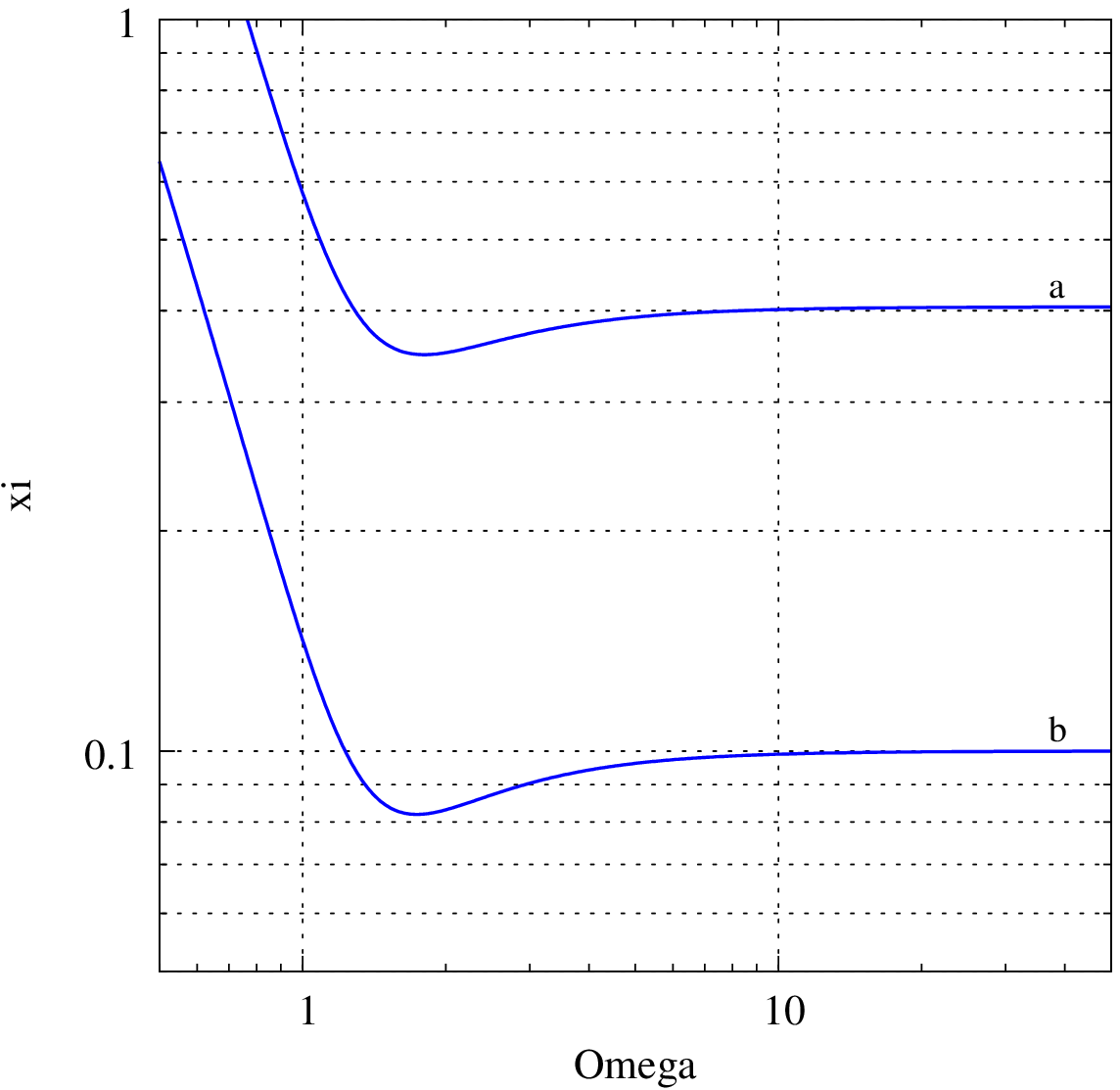}\bigskip
  \caption{Sensitivity dependence on the observation frequency: (a) $l_f=10\,\mathrm{m},\,\Omega_c=2\pi\times500\,\mathrm{s}^{-1},\,W/m\approx50\,\mathrm{W/g}$; (b)  $l_f=4\,\mathrm{km},\,\Omega_c=2\pi\times50\,\mathrm{s}^{-1},\,W/m\approx5\,\mathrm{kW/g}$.}\label{fig:varmeas2}
\end{figure}

In Sec.\ref{app:estimates} two examples of the parameters sets are considered in detail. The first one corresponds to a hypothetical laboratory-scale mechanical QND measurement setup: $l=10\,\mathrm{m},\,\Omega_c=2\pi\times500\,\mathrm{s}^{-1}$, and the second one to the local meter  with long filter cavity: $l=4\,\mathrm{km},\,\Omega_c=2\pi\times50\,\mathrm{s}^{-1}$. The higher  cut-off frequency in the first example compensates partly the short filter cavity length; in the second example, the cut-off frequency is approximately equal to the low-frequency boundary of the Advanced LIGO.

In Fig.\,\ref{fig:varmeas} the factor $\xi$ is plotted as a function of the optical power for these two examples, both for resonant-pumped and optimally-detuned meter cavity. It is easy to see that in the first case the sensitivity $\xi\approx0.4$ can be obtained with specific optical power $W/m\approx50\,\mathrm{W/g}$, and in the second case $\xi\approx0.1$ when $W/m\approx5\,\mathrm{kW/g}$ (see points marked by ``$\otimes$''s on the plot). In Fig.\,\ref{fig:varmeas2} the function
\begin{equation} \xi(\Omega) = \sqrt{\frac{2m\Omega_c^2S(\Omega)}{\hbar}}\end{equation} 
which describes spectral dependence of sensitivity is plotted for these two parameters set.

\section{Mirrors coating and bulk thermal noises}\label{sec:mirrors}

A crude estimate of thermal noises in the mirrors coating and bulk can be obtained by ``scaling down'' the estimates for the Advanced LIGO. The planned mass of the Advanced LIGO mirror is $M=40\,\mathrm{kg}$, and the local mirror mass could be as small as $m\sim 1\,\mathrm{g}$. Therefore, the signal displacement of the local mirror could be $\sqrt{M/m}\sim100$ times larger than the displacement of the heavy Advanced LIGO mirrors. On the other hand, the local mirror radius have to be $r_M/r_m \sim (M/m)^{1/3}\sim30$ times smaller than one of the Advanced LIGO mirrors. It dictates the light beam radius to be proportionally smaller (in order to keep diffraction losses at the same level). It means in turn (see \cite{03a1BrVy}) that\\ (a) the local mirror Brownian noise will increases by the factor of $(r_M/r_m)^{1/2}\sim 5$; \\
(b) the local mirror coating thermoelastic noise will increases by the factor of $r_M/r_m\sim 30$; \\
and (c) the local mirror bulk thermoelastic noise increases by the factor of $(r_M/r_m)^{3/2}\sim 100$.

Therefore, the signal-to-noise ratio increases for the first two noises and remains approximately the same for the third one, which is not considered as the most dangerous (estimates of the paper \cite{03a1BrVy} show that for the Advanced LIGO it is about one and a half order of magnitude smaller than the SQL). 

It is possible to conclude from this simple consideration, that if these noises problem will be solved for the Advanced LIGO, then it will be also solved for small-scale meters. It have to be noted also that, as it was mentioned above, the topology of Fig.\,\ref{fig:rscheme} allows do reduce the mirrors coating noise. 

\section{Conclusion}

It follows from the calculations of this paper that with the best mirrors now available (with losses per bounce $\sim 10^{-5}$), spectral variational measurement can be considered as feasible method of overcoming the Standard Quantum Limit in intracavity readout topologies of future gravitational-wave detectors. Optical losses both in meter cavities and filter cavity limit the sensitivity. However, the influence of former kind of losses can be suppressed by the increase of optical power, and of the latter one influence can be suppressed only by the increase of filter cavity length. 

The sensitivity about ten times better than the Standard Quantum Limit can be achieved by using high, but realistic value of the optical power circulating in the local meter cavity, approximately equal to several kilowatts, small ($m\sim 1\,\mathrm{g}$) local mirror, and long (kilometer-scale) filter cavity. 

A demonstration experiment aimed at beating the Standard Quantum Limit by the factor of $2\div3$ can be performed using much smaller circulating power (about tens of watts ) and relatively short filter cavity (with length $\sim 10\,\mathrm{m})$.

The meter cavities length $l$ does not appear in this paper equations, except the condition (\ref{wb_cond}) and can be chosen according to technological reasons. Probably, the ``sweet point'' is in the $0.1\div1\,\mathrm{m}$ range (see also discussion in \cite{06a1CoChKhOtVyWhMa}).

It should be noted that the necessary optical power can be reduced by injecting squeezed vacuum into the meter cavities. In particular, the recent impressive achievements in preparation of squeezed quantum states in low-frequency ($100\div1000\,\mathrm{Hz}$) band \cite{McKenzie2004, Valbruch2006} allow to reduce all estimates for optical power made in this paper by the factor of $2\div3$.

\acknowledgments

This work was supported by NSF and Caltech grant PHY-0353775 and by Russian government grant NSh-5178.2006.2.

The author is grateful to V.B.Braginsky, Y.Chen, S.Danilishin, K.Danzmann, J.Harms, M.Gorodetsky, H.Luck, R.Schnabel, K.Somiya and S.P.Vyatchanin for stimulating discussions and useful remarks.

The author is grateful also to Y.Chen for hospitality.

\appendix

\section{Notations and approximations}

Additional notations used in the Appendix and not shown in Table\,\ref{tab:notations} are listed in Table \ref{tab:notations2}.
\begin{table}
  \begin{tabular}{|rl|rl|}
    \hline
    $\gamma_I$  & $=\dfrac{cT^2}{4l}$       & $\gamma_{fI}$   & $=\dfrac{cT_f^2}{4l_f}$ \\
    $\gamma_E$  & $=\dfrac{cA^2}{4l}$       & $\gamma_{fE}$   & $=\dfrac{cA_f^2}{4l_f}$ \\
    $\gamma$    & $=\gamma_I+\gamma_E$      & $\gamma_f$      & $=\gamma_{fI}+\gamma_{fE}$ \\ 
    $\delta$    & --- meter cavity detuning & $\delta_f$  & --- filter cavity detuning \\
    $\Delta$    & $=\delta/\gamma$          & & \\
    ${\cal J}$     & $=\dfrac{16\omega_pW}{mc^2T^2(1+{\cal A})(1+\Delta^2)}$ & & \\
    \hline
  \end{tabular}
  \caption{Additional notations not shown in Table\,\ref{tab:notations}.}\label{tab:notations2}
\end{table} 

The field amplitudes are presented as sums of large classical values (denoted by capital roman letters) and small quantum ones (denoted by small roman letter). Only linear in these small quantum fluctuations and in the mirrors displacements $x_I, x_E$ terms are kept. 

It is supposed that optical pumping frequency $\omega_p$ is much larger than all other frequencies, and that the cavity is short:
\begin{align}\label{short_cavity}
  \Omega &\ll \frac{1}{\tau} \,, & \gamma &\ll \frac{1}{\tau} \,,
  & \delta=\omega_p-\omega_o &\ll \frac{1}{\tau} \,.
\end{align}
High (optical range) frequencies are denoted by $\omega$, and low (mechanical-range) ones by $\Omega$. Typically, $\omega=\omega_p+\Omega$.

\section{Single cavity}

\subsection{Field amplitudes}

\begin{figure}

\psfrag{a}[rb][lb]{$\hat {\rm a}$}\psfrag{b}[rb][lb]{$\hat {\rm b}$}
\psfrag{c}[lb][lb]{$\hat {\rm c}$}\psfrag{d}[lb][lb]{$\hat {\rm d}$}
\psfrag{e}[rb][lb]{$\hat {\rm e}$}\psfrag{f}[rb][lb]{$\hat {\rm f}$}
\psfrag{g}[lb][lb]{$\hat {\rm g}$}\psfrag{h}[lb][lb]{$\hat {\rm h}$}
\psfrag{RTI}[cb][lb]{$-R_I,iT$}\psfrag{RTE}[cb][lb]{$-R_E,iA$}
\psfrag{xi}[lb][lb]{$x_I$}\psfrag{xe}[lb][lb]{$x_E$}
\psfrag{L}[ct][lb]{$L=c\tau$}

\begin{center}\includegraphics[width=0.5\textwidth]{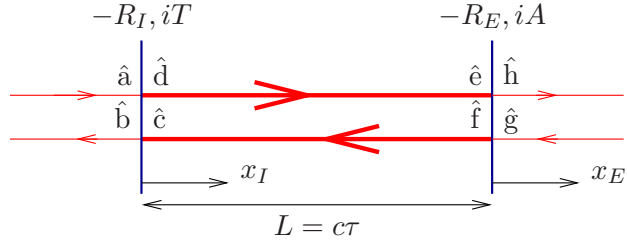}\end{center}

\caption{Single Fabry-Perot cavity}\label{fig:fabry_loss}

\end{figure}

The scheme considered here is shown in Fig.\,\ref{fig:fabry_loss}. $R_I, R_E$ are the mirrors reflectivities. All optical losses are modeled by the end mirror transmittance $A$, so
\begin{equation} R_I^2+T^2 = R_E^2+A^2 = 1\,.\end{equation} 

Equations for the classical field amplitudes are the following:
\begin{subequations}
  \begin{align}
    {\rm B} &= -R_I{\rm A} + iT{\rm C} \,, & {\rm C} &= {\rm F}e^{i\omega_p\tau} \,, \\
    {\rm D} &= -R_I{\rm C} + iT{\rm A} \,, & {\rm E} &= {\rm D}e^{i\omega_p\tau} \,, \\ 
    {\rm F} &= -R_E{\rm E} \,.
  \end{align}
\end{subequations}
It follows from these equations that
\begin{align}\label{FPL_0}
  {\rm C} &= -R_E{\rm E}e^{i\omega_p\tau}\,, & {\rm D} &= {\rm E}e^{-i\omega_p\tau}\,, & 
    {\rm F} &= -R_E{\rm E}\,.
\end{align}

Equations for the quantum field amplitudes are the following:
\begin{subequations}
  \begin{align}
    \hat{\rm b}(\omega) &= -R_I\hat{\rm a}(\omega) + iT\hat{\rm c}(\omega) 
      - \frac{2i\sqrt{\omega\omega_p}}{c}\,{\rm A}R_Ix_I(\Omega)\,,\\
    \hat{\rm c}(\omega) &= \hat{\rm f}(\omega)e^{i\omega\tau} \,, \\
    \hat{\rm d}(\omega) &= -R_I\hat{\rm c}(\omega) + i\hat s_I(\omega)\,, \\
    \hat{\rm e}(\omega) &= \hat{\rm d}(\omega)e^{i\omega\tau} \,, \\
    \hat{\rm f}(\omega) &= -R_E\hat{\rm e}(\omega) + i\hat s_E(\omega)\,,
  \end{align}
\end{subequations}
where
\begin{subequations}\label{FPL_IE}
  \begin{align}
    \hat s_I(\omega) &= T\hat{\rm a}(\omega) 
      + \frac{2\sqrt{\omega\omega_p}}{c}\,{\rm C}R_Ix_I(\Omega) \,, \\
    \hat s_E(\omega) &= A\hat{\rm g}(\omega) 
      - \frac{2\sqrt{\omega\omega_p}}{c}\,{\rm E}R_Ex_E(\Omega) \,.
  \end{align}
\end{subequations}
Solution of these equations is the following:
\begin{subequations}\label{FPL_1}
  \begin{multline}
    \hat{\rm b}(\omega) = \frac{1}{\det(\omega)}\bigl[
        (R_Ee^{2i\omega\tau} - R_I)\hat{\rm a}(\omega) - T\hat s_E(\omega)e^{i\omega\tau}
        + \frac{2\sqrt{\omega\omega_p}}{c}\,{\rm C}TR_ER_Ix_I(\Omega)e^{2i\omega\tau}
      \bigr] \\
      - \frac{2i\sqrt{\omega\omega_p}}{c}\,{\rm A}R_Ix_I(\Omega) \,, \label{FPL_1(b)} 
  \end{multline}\\[-2ex]
  \begin{align}
    \det(\omega)\hat{\rm c}(\omega) &= -iR_E\hat s_I(\omega)e^{2i\omega\tau} 
      + i\hat s_E(\omega)e^{i\omega\tau} \,, \\
    \det(\omega)\hat{\rm d}(\omega) &= i\hat s_I(\omega) - iR_I\hat s_E(\omega)e^{i\omega\tau} \,, \\
    \det(\omega)\hat{\rm e}(\omega) &= i\hat s_I(\omega)e^{i\omega\tau} 
      - iR_I\hat s_E(\omega)e^{2i\omega\tau} \,, \\
    \det(\omega)\hat{\rm f}(\omega) &= -iR_E\hat s_I(\omega)e^{i\omega\tau} + i\hat s_E(\omega) \,,
  \end{align}
\end{subequations}
where
\begin{equation} \det(\omega) = 1-R_IR_Ee^{2i\omega\tau} \,. \end{equation} 

Approximation (\ref{short_cavity}) allows to simplify Eqs.\,(\ref{FPL_0}, \ref{FPL_1}):
\begin{equation}\label{FPL_0_NB}
  -{\rm C} = {\rm D} = {\rm E} = -{\rm F} \,,
\end{equation} 
and
\begin{subequations}\label{FPL_1_NB}
  \begin{gather}
    \hat{\rm b}(\omega) = \frac{1}{\ell(\Omega)}\left[
        [\ell^*(\Omega)-2\gamma_E]\hat{\rm a}(\omega)
        - 2\sqrt{\gamma_I\gamma_E}\,\hat{\rm g}(\omega) 
        + \frac{2\omega_p}{c}\,\sqrt{\frac{\gamma_I}{\tau}}\,{\rm E}x(\Omega)
      \right] \,, \label{FPL_1_NB(b)} \\
    -\hat{\rm c}(\omega) = \hat{\rm d}(\omega) = \hat{\rm e}(\omega) = -\hat{\rm f}(\omega)
      = \frac{1}{\ell(\Omega)}\left[
          i\sqrt{\frac{\gamma_I}{\tau}}\,\hat{\rm a}(\omega) 
          - i\sqrt{\frac{\gamma_E}{\tau}}\,\hat{\rm g}(\omega) + \frac{i\omega_p{\rm E}x(\Omega)}{l}
        \right] \,, \label{FPL_1_NB_e} 
  \end{gather}
\end{subequations}
where
\begin{gather}
  \ell(\Omega) = \gamma-i(\delta+\Omega) \,, \\
  x(\Omega) = x_E(\Omega)-x_I(\Omega) \,.
\end{gather}

It is convenient to introduce ``rotated'' input noises:
\begin{subequations}
  \begin{align}
    \hat{\rm a}_{\rm new}(\omega) &= \frac{\ell^*(\Omega)}{\ell(\Omega)}\,\frac{
        \sqrt{\gamma_I}\,\hat{\rm a}(\omega) - \sqrt{\gamma_E}\,\hat{\rm g}(\omega)
      }{\sqrt{\gamma}}\,, \\
    \hat{\rm g}_{\rm new}(\omega) &= \frac{
        \sqrt{\gamma_I}\,\hat{\rm g}(\omega) + \sqrt{\gamma_E}\,\hat{\rm a}(\omega)
      }{\sqrt{\gamma}}\,.
  \end{align}
\end{subequations}
As $\hat{\rm a}$ and $\hat{\rm g}$ describe independent zero-point fluctuations, this redefinition does not affect the end result. Using these noises, we obtain (subscript ``new'' is omitted for simplicity):
\begin{subequations}\label{FPL_1_N2}
  \begin{gather}
    \hat{\rm b}(\omega) = \frac{
        \sqrt{\gamma_I}\,\hat{\rm a}(\omega) - \sqrt{\gamma_E}\,\hat{\rm g}(\omega)
      }{\sqrt{\gamma}}
      + \sqrt{\frac{\gamma_I}{\tau}}\,\frac{2\omega_p{\rm E}x(\Omega)}{c\ell(\Omega)}\,,
      \label{FPL_1_NB2(a)} \\
    -\hat{\rm c}(\omega) = \hat{\rm d}(\omega) = \hat{\rm e}(\omega) = -\hat{\rm f}(\omega)
      = i\sqrt{\frac{\gamma}{\tau}}\,\frac{\hat{\rm a}(\Omega)}{\ell^*(\Omega)} 
        + \frac{i\omega_p{\rm E}x(\Omega)}{l\ell(\Omega)}\,. \label{FPL_1_NB2(e)} 
  \end{gather}
\end{subequations}

\subsection{Pondermotive forces}

In the spectral representation, pondermotive forces acting on the mirrors are equal to:
\begin{equation}
  F_{I,E}(\Omega) = {\cal F}_{I,E}(\Omega) + {\cal F}_{I,E}^*(-\Omega) \,,
\end{equation}
where
\begin{subequations}
  \begin{align}
    {\cal F}_I(\Omega) &= \frac{\hbar\sqrt{\omega\omega_p}}{c}\,\Bigl[
        {\rm A}^*\hat{\rm a}(\omega) + {\rm B}^*\hat{\rm b}(\omega) 
        - {\rm C}^*\hat{\rm c}(\omega) - {\rm D}^*\hat{\rm d}(\omega) 
      \Bigr] \,, \\
    {\cal F}_E(\Omega) &= \frac{\hbar\sqrt{\omega\omega_p}}{c}\,\Bigl[
        {\rm E}^*\hat{\rm e}(\omega) + {\rm F}^*\hat{\rm f}(\omega) - {\rm H}^*\hat{\rm h}(\omega) 
      \Bigr] \,.
  \end{align}
\end{subequations}
Approximation (\ref{short_cavity}) gives that
\begin{gather}
  -{\cal F}_I(\Omega) = {\cal F}_E(\Omega) 
  = {\cal F}(\Omega) = \frac{2\hbar\omega_p{\rm E}^*\hat{\rm e}(\omega)}{c} \,, \\
  -F_I(\Omega) = F_E(\Omega) = F(\Omega) = \hat F_{\rm fluct}(\Omega) - K(\Omega)x(\Omega)
  \label{F_pond} \,,
\end{gather} 
where
\begin{equation}\label{FPL_F_fl}
  \hat F_{\rm fluct}(\Omega) 
  = \frac{2i\hbar\omega_p}{c}\sqrt{\dfrac{\gamma}{\tau}}\left[
      \dfrac{{\rm E}^*\hat{\rm a}(\omega)}{\ell^*(\Omega)}
      - \dfrac{{\rm E}\hat{\rm a}^+(-\omega)}{\ell(-\Omega)}
    \right] 
\end{equation}
is the fluctuational (back-action) force and
\begin{equation}
  K(\Omega) = \frac{4\hbar\omega_p^2|{\rm E}|^2\delta}{cl\ell(\Omega)\ell^*(-\Omega)}
  = \frac{4\omega_pW\delta}{cl\ell(\Omega)\ell^*(-\Omega)}
\end{equation}
is the pondermotive rigidity.

Consider now the mechanical equations of motion:
\begin{subequations}\label{mech_eq}
  \begin{align}
    -m_I\Omega^2x_I(\Omega) &= F_I(\Omega) \,, \\
    -m_E\Omega^2x_E(\Omega) &= F_E(\Omega) - m_E\Omega^2x_{\rm signal}(\Omega)\,,
  \end{align}
\end{subequations}
where $-m_E\Omega^2x_{\rm signal}(\Omega)$ is the signal force (in general, $x_{\rm signal}(\Omega)$ is {\em not} equal to the signal displacement). Eqs.\,(\ref{F_pond}, \ref{mech_eq}) give that:
\begin{equation}\label{FPL_x}
  x(\Omega) = \frac{F_{\rm fluct}(\Omega)-m\Omega^2x_{\rm signal}(\Omega)}{K(\Omega)-m\Omega^2}\,,
\end{equation}
where
\begin{equation}
  m = \left(\frac{1}{m_I} + \frac{1}{m_E}\right)^{-1} 
\end{equation}
is the reduced mass.

\subsection{The sum noise spectral density}

Substitution of Eq.\,(\ref{FPL_x}) in Eq.\,(\ref{FPL_1_NB2(a)}) gives:
\begin{equation}\label{FPL_b_out}
  \hat{\rm b}(\omega) = \sqrt{\frac{\gamma_I}{\gamma}}\left[
    \alpha(\Omega)\hat{\rm a}(\Omega) + \beta(\Omega)\hat{\rm a}^+(-\omega)
  \right]
  - \sqrt{\frac{\gamma_E}{\gamma}}\,\hat{\rm g}(\omega) + X(\Omega)x_{\rm signal}(\Omega) \,,
\end{equation}
where
\begin{subequations}\label{FPL_alpha_beta_X}
  \begin{align}
    \alpha(\Omega) 
      &=1 + \frac{4i\omega_pW\gamma}{cl[K(\Omega)-m\Omega^2]|\ell(\Omega)|^2} \,, \\
    \beta(\Omega) 
      &=-\frac{4i\omega_pW\gamma e^{2i\arg E}}{cl[K(\Omega)-m\Omega^2]\ell(\Omega)\ell(-\Omega)} \,, \\
    X(\Omega) &= \sqrt{\frac{4\omega_pW\gamma_I}{\hbar cl}}\,
      \frac{-m\Omega^2}{K(\Omega)-m\Omega^2}\,\frac{e^{i\arg E}}{\ell(\Omega)} \,.
  \end{align}
\end{subequations}
The homodyne detector output for the field (\ref{FPL_b_out}) is equal to:
\begin{equation}
  \hat y(t) 
   = \int_0^\infty\hat{\rm b}(\omega)e^{-i\Omega t+i\phi_{\rm LO}}\,\frac{d\omega}{2\pi} + \hc\,,
\end{equation}
where $\phi_{LO}$ is the local oscillator phase. Corresponding sum noise spectral density is equal to:
\begin{equation}\label{FPL_Ssum_1}
  S(\Omega) = \frac{
    \dfrac{\gamma_I}{\gamma}
      \left|\alpha(\Omega)e^{i\phi_{\rm LO}} + \beta^*(-\Omega)e^{-i\phi_{\rm LO}}\right|^2
    + \dfrac{\gamma_E}{\gamma}
  }{\left|X(\Omega)e^{i\phi_{\rm LO}} + X^*(-\Omega)e^{-i\phi_{\rm LO}}\right|^2} \,.
\end{equation}

In the particular case of wide band cavity [see Eq.(\ref{wb_cond})],
\begin{subequations}\label{FPLs_alpha_beta_X}
  \begin{gather}
    K = m\Omega_0^2 = m{\cal J}\Delta \,, \label{FPLs_alpha_beta_X(a)} \\
    \alpha(\Omega) = 1 + \frac{i{\cal J}}{\Omega_0^2-\Omega^2}\,, \\
    \beta(\Omega) = -\frac{i{\cal J}}{\Omega_0^2-\Omega^2}\,\exp\left[2i(\arg E+\arctan\Delta)\right]\,,\\
    X(\Omega) = \sqrt{\frac{m{\cal J}}{\hbar(1+{\cal A})}}\,\frac{-\Omega^2}{\Omega_0^2-\Omega^2}\,
      \exp\left[i(\arg E+\arctan\Delta)\right]\,, 
  \end{gather}
\end{subequations}
and
\begin{equation}\label{S_sum}
  S(\Omega) = \frac{m^2(\Omega_0^2-\Omega^2)S_x + 2m(\Omega_0^2-\Omega^2)S_{xF} + S_F}{m^2\Omega^4}
  = S_x^{\rm eff} - \frac{2S_{xF}^{\rm eff}}{m\Omega^2} + \frac{S_F^{\rm eff}}{m^2\Omega^4} \,,
\end{equation} 
where
\begin{subequations}\label{Sx_SF_phi_const}
  \begin{align}
    S_x &= \frac{\hbar(1+{\cal A})}{4m{\cal J}\cos^2\phi}
      = \frac{\hbar c^2T^2(1+\Delta^2)(1+{\cal A})^2}{64\omega_pW\cos^2\phi}\,,
        \label{Sx_SF_phi_const(a)}\\ 
    S_F &= \hbar m{\cal J} = \frac{16\hbar\omega_pW}{c^2T^2(1+\Delta^2)(1+{\cal A})} \,, \\
    S_{xF} &= \frac{\hbar}{2}\,\tan\phi \,, 
  \end{align}
\end{subequations}
\begin{subequations}\label{Sx_SF_eff}
  \begin{align}
    S_x^{\rm eff} &= S_x\,, \\
    S_F^{\rm eff} &= \frac{\hbar m{\cal J}\cos^2\phi}{1+{\cal A}}\left\{
        1 + \frac{{\cal A}}{\cos^2\phi} 
        + \left[\tan\phi + \frac{\Delta(1+{\cal A})}{2\cos^2\phi}\right]^2
      \right\}\,, \\
    S_{xF}^{\rm eff} &= \frac{\hbar}{2}\left[\tan\phi + \frac{\Delta(1+{\cal A})}{2\cos^2\phi}\right] \,,
  \end{align}
\end{subequations}
and
\begin{equation}
  \phi = \phi_{\rm LO} + \arg E+\arctan\Delta \,.
\end{equation} 

If ${\cal A}=0$ (no optical losses) and $\Delta=0\ \Rightarrow\ \Omega_0=0$ (resonant pumping), then
\begin{subequations}\label{Sx_SF_res}
  \begin{align}
    S_x^{\rm eff} = S_x &= \frac{\hbar c^2T^2}{64\omega_pW\cos^2\phi}\,,\\
    S_F^{\rm eff} = S_F &= \frac{16\hbar\omega_pW}{c^2T^2} \,, \\
    S_{xF}^{\rm eff} = S_{xF} &= \frac{\hbar}{2}\,\tan\phi \,.
  \end{align}
\end{subequations}
If also $\phi=0\ \Rightarrow\ S_{xF}=0$ (no noises correlation), then 
\begin{align}\label{Sx_SF_0}
  S_x &= \frac{\hbar c^2T^2}{64\omega_pW} \,, & S_F &= \frac{16\hbar\omega_pW}{c^2T^2} \,.
\end{align}

\section{Balanced scheme}\label{app:corbitt_scheme}.

\begin{figure*}

\psfrag{a}[rb][lb]{$\hat {\rm a}_1$}\psfrag{b}[rb][lb]{$\hat {\rm b}_1$}
\psfrag{c}[lb][lb]{$\hat {\rm c}_1$}\psfrag{d}[lb][lb]{$\hat {\rm d}_1$}
\psfrag{e}[rb][lb]{$\hat {\rm e}_1$}\psfrag{f}[rb][lb]{$\hat {\rm f}_1$}
\psfrag{g}[lb][lb]{$\hat {\rm g}_1$}\psfrag{h}[lb][lb]{$\hat {\rm h}_1$}
\psfrag{a2}[lb][lb]{$\hat {\rm a}_2$}\psfrag{b2}[lb][lb]{$\hat {\rm b}_2$}
\psfrag{c2}[rb][lb]{$\hat {\rm c}_2$}\psfrag{d2}[rb][lb]{$\hat {\rm d}_2$}
\psfrag{e2}[lb][lb]{$\hat {\rm e}_2$}\psfrag{f2}[lb][lb]{$\hat {\rm f}_2$}
\psfrag{g2}[rb][lb]{$\hat {\rm g}_2$}\psfrag{h2}[rb][lb]{$\hat {\rm h}_2$}
\psfrag{RTI}[cb][lb]{$-R_I,iT$}\psfrag{RTE}[cb][lb]{$-R_E,iA$}
\psfrag{xi}[lb][lb]{$x_{I1}$}\psfrag{xe}[rt][lb]{$x_{E1}$}
\psfrag{xi2}[rb][lb]{$x_{I2}$}\psfrag{xe2}[lt][lb]{$x_{E2}$}
\psfrag{L}[ct][lb]{$L=c\tau$}
\psfrag{AT}[cb][lb]{$A_0,T_0$}

\begin{center}\includegraphics[width=0.6\textwidth]{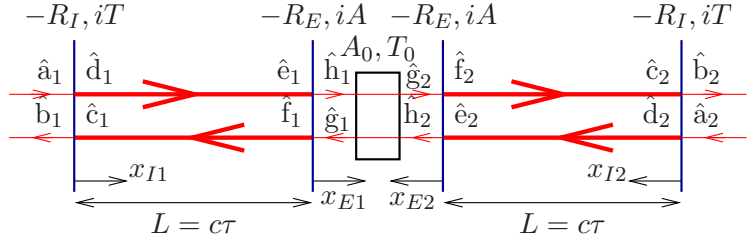}\end{center}

\caption{Balanced scheme}\label{fig:corbitt_loss}

\end{figure*}

Consider a combination of two Fabry-Perot cavities, see Fig.\,\ref{fig:corbitt_loss}. The central block has transmittance $T_0$ and absorption $A_0$, 
\begin{equation}
  |T_0^2| + |A_0|^2 = 1 \,,
\end{equation} 
and models losses in the end mirrors substrate. For simplicity, losses in the mirrors coating are not taken into consideration because the goal of this Appendix is just to show the equivalence between balanced scheme with partly transparent end mirrors and single Fabry-Perot cavity.

The following symmetric and anti-symmetric modes will be used:
\begin{equation} {\rm a}_\pm = \frac{{\rm a}_1 \pm {\rm a}_2}{\sqrt{2}} \,, \end{equation} 
and so on for all other field amplitudes. It is well known that in the Michelson---Fabry-Perot topologies, including the one shown in Fig.\,\ref{fig:rscheme}, the pumping power input is coupled with the symmetric (``$+$'') mode and the signal (dark port) output is coupled with the anti-symmetric (``$-$'' in our notations) mode. Therefore, classical field amplitudes will be considered for the ``$+$'' mode only (and the ones for ``$-$'' mode will be set equal to zero), quantum ones for the ``$-$'' mode only, and subscripts ``$+$'' and ``$-$'' will be omitted for simplicity.

Equations for the classical field amplitudes are the following:
\begin{subequations}
  \begin{align}
    {\rm B}_{1,2} &= -R_I{\rm A}_{1,2} + iT{\rm C}_{1,2} \,, & 
    {\rm C}_{1,2} &= {\rm F}_{1,2}e^{i\omega_p\tau} \,, \\
    {\rm D}_{1,2} &= -R_I{\rm C}_{1,2} + iT{\rm A}_{1,2} \,, &
    {\rm E}_{1,2} &= {\rm D}_{1,2}e^{i\omega_p\tau}\,, \\
    {\rm F}_{1,2} &= -R_E{\rm E}_{1,2} + iA{\rm G}_{1,2}\,, &
    {\rm G}_{1,2} &= T_0{\rm H}_{2,1} \,, \\
    {\rm H}_{1,2} &= -R_E{\rm G}_{1,2} + iA{\rm E}_{1,2} \,.
  \end{align}
\end{subequations}
Therefore,
\begin{subequations}\label{DFPL_EQ_0}
  \begin{align}
    {\rm B} &= -R_I{\rm A} + iT{\rm C} \,, & {\rm C} &= {\rm F}e^{i\omega_p\tau} \,,
       \label{DFPL_EQ_0(bc)} \\
    {\rm D} &= -R_I{\rm C} + iT{\rm A} \,, & {\rm E} &= {\rm D}e^{i\omega_p\tau} \,,
       \label{DFPL_EQ_0(de)} \\
    {\rm F} &= -R_E{\rm E} + iA{\rm G} \,, & {\rm G} &= T_0{\rm H} \,, \label{DFPL_EQ_0(fg)} \\
    {\rm H} &= -R_E{\rm G} + iA{\rm E} \,. \label{DFPL_EQ_0(h)}
  \end{align}
\end{subequations}
It follows from Eqs.\,(\ref{DFPL_EQ_0(fg)}, \ref{DFPL_EQ_0(h)}) that
\begin{equation}\label{DFPL_F_0} {\rm F} = -\frac{T_0+R_E}{1+T_0R_E}\,{\rm E} \,. \end{equation} 

Equations for the quantum field amplitudes are the following:
\begin{subequations}
  \begin{align}
    \hat{\rm b}_{1,2}(\omega) &= -R_I\hat{\rm a}_{1,2}(\omega) + iT\hat{\rm c}_{1,2}(\omega) 
      - \frac{2i\sqrt{\omega\omega_p}}{c}\,{\rm A}_{1,2}R_Ix_{I\,1,2}(\Omega) \,, \\
    \hat{\rm c}_{1,2}(\omega) &= \hat{\rm f}_{1,2}(\omega)e^{i\omega\tau} \,, \\
    \hat{\rm d}_{1,2}(\omega) &= -R_I\hat{\rm c}_{1,2}(\omega) + i\hat s_{I\,1,2}(\omega)\,, \\
    \hat{\rm e}_{1,2}(\omega) &= \hat{\rm d}_{1,2}(\omega)e^{i\omega\tau} \,, \\
    \hat{\rm f}_{1,2}(\omega) &= -R_E\hat{\rm e}_{1,2}(\omega) + i\hat s_{E\,1,2}(\omega)\,, \\
    \hat{\rm g}_{1,2}(\omega) &= T_0\hat{\rm h}_{2,1}(\omega) + A_0\hat{\rm n}_{1,2}(\omega)\,, \\
    \hat{\rm h}_{1,2}(\omega) &= -R_E\hat{\rm g}_{1,2}(\omega) + iA\hat{\rm e}_{1,2}(\omega)
      + \frac{2i\sqrt{\omega\omega_p}}{c}\,{\rm G}_{1,2}R_Ex_{E\,1,2}(\Omega) \,,
  \end{align}
\end{subequations}
where
\begin{gather}
  \hat s_{I\,1,2}(\omega) = T\hat{\rm a}_{1,2}(\omega) 
    + \frac{2\sqrt{\omega\omega_p}}{c}\,{\rm C}_{1,2}R_Ix_{I\,1,2}(\Omega) \,, \\
  \hat s_{E\,1,2}(\omega) = A\hat{\rm g}_{1,2}(\omega) 
    - \frac{2\sqrt{\omega\omega_p}}{c}\,{\rm E}_{1,2}R_Ex_{E\,1,2}(\Omega) \,,
\end{gather}
and $\hat{\rm n}_{1,2}(\omega)$ are noises created by losses in the middle block. Therefore,
\begin{subequations}
  \begin{align}
    \hat{\rm b}(\omega) &= -R_I\hat{\rm a}(\omega) + iT\hat{\rm c}(\omega) 
      - \frac{2i\sqrt{\omega\omega_p}}{c}\,{\rm A}R_Ix_I(\Omega) \,, \label{DFPL_EQ_1(b)} \\
    \hat{\rm c}(\omega) &= \hat{\rm f}(\omega)e^{i\omega\tau} \,,  \\
    \hat{\rm d}(\omega) &= -R_I\hat{\rm c}(\omega) + i\hat s_I(\omega)\,, \\
    \hat{\rm e}(\omega) &= \hat{\rm d}(\omega)e^{i\omega\tau} \,, \label{DFPL_EQ_1(e)} \\
    \hat{\rm f}(\omega) &= -R_E\hat{\rm e}(\omega) + i\hat s_E(\omega)\,, \label{DFPL_EQ_1(f)}  \\
    \hat{\rm g}(\omega) &= - T_0\hat{\rm h}(\omega) + A_0\hat{\rm n}(\omega)\,, \\
    \hat{\rm h}(\omega) &= -R_E\hat{\rm g}(\omega) + iA\hat{\rm e}(\omega) 
      + \frac{2i\sqrt{\omega\omega_p}}{c}\,{\rm G}R_Ex_E(\Omega)\,, \label{DFPL_EQ_1(h)}
  \end{align}
\end{subequations}
where
\begin{subequations}
  \begin{align}
    \hat s_I(\omega) &= T\hat{\rm a}(\omega) 
      + \frac{2\sqrt{\omega\omega_p}}{c}\,{\rm C}R_Ix_I(\Omega) \,, \\
    \hat s_E(\omega) &= A\hat{\rm g}(\omega) 
      - \frac{2\sqrt{\omega\omega_p}}{c}\,{\rm E}R_Ex_E(\Omega)\,, 
  \end{align}
\end{subequations}
\begin{subequations}
  \begin{align}
    x_I(\Omega) &= \frac{x_{I1}(\Omega) - x_{I2}(\Omega)}{2} \,, \\
    x_E(\Omega) &= \frac{x_{E1}(\Omega) - x_{E2}(\Omega)}{2} \,.
  \end{align}
\end{subequations}
It follows from Eqs.\,(\ref{DFPL_EQ_1(f)}---\ref{DFPL_EQ_1(h)}) that
\begin{equation}
  \hat{\rm f}(\omega) = \frac{(T_0-R_E)\hat{\rm e}(\omega) + iA_0A\hat{\rm n}(\Omega)}{1-T_0R_E}
    + \frac{2i\sqrt{\omega\omega_p}}{c}\,\frac{T_0^2-1}{1-T_0^2R_E}\,{\rm E}R_Ex_E(\Omega) \,.
   \label{DFPL_F_1}
\end{equation} 

Suppose now that $T_0 = i|T_0|$. This condition corresponds to the anti-resonance tuning of the cavity formed by two end mirrors. Suppose also that $A\ll 1$, $|A_0|\ll 1$. In this case,
\begin{align}
  {\rm F} &\approx -R_{\rm eff}{\rm E}e^{iA^2/2} \,, \label{DFPL_0_eff} \\
  \hat{\rm f}(\omega) &\approx -R_{\rm eff}\hat{\rm e}(\omega)e^{-iA^2/2} 
    + iT_{\rm eff}\hat{\rm n}(\Omega)e^{i\pi/4}
    - \frac{2i\sqrt{\omega\omega_p}}{c}\,{\rm E}\sqrt{R_E}\,x_E(\Omega) \label{DFPL_1_eff} \,,
\end{align}
where
\begin{align}\label{DFPL_RAff}
  R_{\rm eff} &= \sqrt{1-T_{\rm eff}^2} \,, & T_{\rm eff} &= \frac{|A_0|A}{\sqrt{2}} \,.
\end{align}
It is easy to see that the equations set (\ref{DFPL_EQ_0(bc)}, \ref{DFPL_EQ_0(de)}, \ref{DFPL_0_eff}) is virtually identical to the equations set (\ref{FPL_0}), and the set (\ref{DFPL_EQ_1(b)}---\ref{DFPL_EQ_1(e)}, \ref{DFPL_1_eff}) to (\ref{FPL_1}), with the end mirror reflectivity and transmittance replaced by effective values (\ref{DFPL_RAff}). The only significant difference is the phase shifts $\pm A^2/2$, which lead to the detuning between symmetric and anti-symmetric modes equal to 
\begin{equation}
  \delta_0 = \frac{cA^2}{2l} 
\end{equation} 
(compare with the cavities bandwidth $\gamma\approx cT^2/4l$). For example, if signal (symmetric) mode is tuned in resonance, then power (anti-symmetric) mode will be detuned by $\delta_0$. However, this detuning can be compensated  (if necessary) by a power recycling mirror (see Fig.\,\ref{fig:rscheme}). 

\section{The variational measurement scheme}\label{app:var_corr}

\subsection{The sum noise spectral density}

Input-output relation for the filter cavity has the form similar to one for the meter cavity (\ref{FPL_1_NB(b)}) with the only difference that there is no pumping power in the filter cavity:
\begin{equation}\label{KLMTV_post_a}
  \hat{\rm b}_f(\omega) = R_f(\Omega)\hat{\rm a}_f(\omega) + T_f(\Omega)\hat{\rm g}_f(\omega) \,,
\end{equation} 
where
\begin{align}
  R_f(\Omega) &= \frac{\gamma_{fI}-\gamma_{fE}+i(\delta_f+\Omega)}{\gamma_f-i(\delta_f+\Omega)}\,, &
  T_f(\Omega) &= -\frac{2\sqrt{\gamma_{fI}\gamma_{fE}}}{\gamma_f-i(\delta_f+\Omega)} \,.
\end{align}
The input field here is the output field of the meter cavity:
\begin{equation} \hat{\rm a}_f(\omega) = \hat{\rm b}(\omega) \,, \end{equation} 
see Eq.\,(\ref{FPL_1_NB2(a)}). Therefore,
\begin{multline}
  \hat{\rm b}_f(\omega) = R_f(\Omega)\left\{
    \frac{
      \sqrt{\gamma_I}
        \left[\alpha(\Omega)\hat{\rm a}(\omega) + \beta(\Omega)\hat{\rm a}^+(-\omega)\right]
      - \sqrt{\gamma_E}\,\hat{\rm g}(\omega)
    }{\sqrt{\gamma}}  
    + X(\Omega)x_{\rm signal}(\Omega)
  \right\} \\
  + T_f(\Omega)\hat{\rm g}_f(\Omega)
\end{multline} 
The corresponding sum noise spectral density is equal to [compare with Eq.\,(\ref{FPL_Ssum_1})]:
\begin{multline}\label{KLMTV_S_orig}
  S(\Omega) = \frac{1}{
    \left|R_f(\Omega)X(\Omega)e^{i\phi_{\rm LO}}
    + R_f^*(-\Omega)X^*(-\Omega)e^{-i\phi_{\rm LO}}\right|^2
  } \\
  \times\biggl(1 
    + \dfrac{\gamma_I}{\gamma}\biggl\{
        \dfrac{|R_f(\Omega)|^2 + |R_f(-\Omega)|^2}{2}\,
          \left[|\alpha(\Omega)|^2 + |\beta(\Omega)|^2 -1\right] \\
        + 2\Re\left[R_f(\Omega)R_f(-\Omega)\alpha(\Omega)\beta(-\Omega)e^{2i\phi_{\rm LO}}\right]
      \biggr\}
  \biggr) \,.
\end{multline}

Suppose that losses in filter cavity are small:
\begin{equation}\label{KLMTVs_cond}
  \gamma_{fE} \ll \sqrt{\gamma_{fI}^2 + (\delta_f\pm\Omega)^2} \,.
\end{equation}
In this case,
\begin{equation}
  R_f(\Omega) \approx \left[1-\frac{2\gamma_{fI}\gamma_{fE}}{\gamma_{fI}^2+(\delta_f+\Omega)^2}\right]
    \exp\left(2i\arctan\frac{\delta_f+\Omega}{\gamma_{f0}}\right) \,,
\end{equation}
and
\begin{subequations}
  \begin{align}
    \frac{|R_f(\Omega)|^2 + |R_f(-\Omega)|^2}{2} &\approx 1-{\cal A}_f(\Omega) \,,  \\
    R_f(\Omega)R_f(-\Omega) &\approx [1-{\cal A}_f(\Omega)]e^{2i\phi_f(\Omega)} \,,
  \end{align}
\end{subequations}
where
\begin{subequations}
  \begin{gather}
    {\cal A}_f(\Omega) = \frac{4\gamma_{fI}\gamma_{fE}(\Omega^2+\gamma_{fI}^2+\delta_f^2)}
      {\Omega^4 + 2(\gamma_{fI}^2-\delta_f^2)\Omega^2 + (\gamma_{fI}^2+\delta_f^2)^2}
      \label{KLMTVs_T2}\,, \\
    \phi_f(\Omega) = \arctan\frac{2\gamma_{fI}\delta}{\gamma_{fI}^2 - \delta_f^2+\Omega^2} \,.
  \end{gather}
\end{subequations}
Suppose also that condition (\ref{wb_cond}) is fulfilled [see also Eqs.(\ref{FPLs_alpha_beta_X})]. In this case,
\begin{equation}\label{KLMTV_S_sum}
  S(\Omega) = \frac{
    m^2(\Omega_0^2-\Omega^2)^2S_x(\Omega) + 2m(\Omega_0^2-\Omega^2)S_{xF}(\Omega) + S_F
  }{m^2\Omega^4} \,,
\end{equation} 
where
\begin{subequations}\label{Sx_SF_phi_var}
  \begin{align}
    S_x &= \frac{\hbar[1+{\cal A}_\Sigma(\Omega)]}{4m{\cal J}\cos^2\phi_\Sigma(\Omega)} \,, \\
    S_F &= \hbar m{\cal J} \,, \\
    S_{xF} &= \frac{\hbar}{2}\,\tan\phi_\Sigma(\Omega)
  \end{align}
\end{subequations}
[compare with Eqs.\,(\ref{S_sum}, \ref{Sx_SF_phi_const})], and
\begin{subequations}
  \begin{align}
    {\cal A}_\Sigma(\Omega) &= {\cal A} + {\cal A}_f(\Omega) \,,\\  
    \phi_\Sigma(\Omega) &= \phi + \phi_f(\Omega) \,.
  \end{align}
\end{subequations}

\subsection{Optimization}

The optimal function $\phi_\Sigma(\Omega)$ which minimize the sum noise spectral density (\ref{KLMTV_S_sum}) is equal to:
\begin{equation}\label{KLMTVs_phi_raw}
  \tan\phi_\Sigma(\Omega) = \frac{{\cal J}}{(\Omega^2-\Omega_0^2)[1 + {\cal A}_\Sigma(\Omega)]}\,.
\end{equation}
Due to the dependence of ${\cal A}_\Sigma$ on $\Omega$, condition (\ref{KLMTVs_phi_raw}) can not be satisfied for all frequencies. Suppose instead that
\begin{equation}\label{KLMTVs_phi_opt}
  \tan\phi_\Sigma(\Omega) = \frac{{\cal J}}{(\Omega^2-\Omega_0^2)(1 + \tilde{\cal A}_\Sigma)}\,,
\end{equation}
where $\tilde{\cal A}_\Sigma$ is some constant which approximates ${\cal A}_\Sigma(\Omega)$. In this case, \begin{equation}\label{KLMTVs_SF_opt}
  S(\Omega) = \frac{\hbar}{m\Omega^4}\left\{
    \frac{(\Omega^2-\Omega_0^2)^2[1 + {\cal A}_\Sigma(\Omega)]}{4{\cal J}}
    + \frac{{\cal J}[{\cal A}_\Sigma(\Omega) + \tilde{\cal A}_\Sigma^2]}{(1+\tilde{\cal A}_\Sigma)^2}
\right\}
\end{equation}
Omitting small in ${\cal A}_\Sigma(\Omega)$ and $\tilde{\cal A}_\Sigma$ terms, we obtain:
\begin{equation}\label{KLMTV_S}
  S(\Omega) = \frac{\hbar}{m\Omega^4}
    \left[\frac{(\Omega^2-\Omega_0^2)^2}{4{\cal J}} + {\cal J}{\cal A}_\Sigma(\Omega)\right] \,.
\end{equation} 
Hence it is possible to set $\tilde{\cal A}_\Sigma=0$, as it does not affect the end result.

In this case, equations (\ref{KLMTVs_T2}) and (\ref{KLMTVs_phi_opt}) give:
\begin{subequations}\label{KLMTVs_parms0}
  \begin{gather}
    \phi = 0 \,, \\
    \gamma_{fI}^2 = \frac{{\cal J}}{2}\left(\sqrt{4+\Delta^2} - \Delta\right) \,,
      \label{KLMTVs_parms0(b)} \\
    \delta_f^2 = \frac{{\cal J}}{2}\left(\sqrt{4+\Delta^2} + \Delta\right) \,, \label{KLMTVs_parms0(c)} \\
    {\cal A}_f(\Omega) = \frac{4\gamma_{fI}\gamma_{fE}\left(\Omega^2+{\cal J}\sqrt{4+\Delta^2}\right)}
      {({\cal J}\Delta-\Omega^2)^2 + 4{\cal J}^2} \,.
  \end{gather}
\end{subequations}
Due to the factor ${\cal A}_\Sigma(\Omega)$, spectral density (\ref{KLMTV_S}) has more sophisticated frequency dependence than ones of (\ref{S_SQL_1}, \ref{S_SQL_corr}). However, in the most interesting case $\Omega\ll{\cal J}$ it is possible to assume that
\begin{subequations}\label{KLMTVs_A_f_0}
  \begin{gather}
    {\cal A}_f(\Omega) \approx {\cal A}_f(0) = \frac{4\gamma_{fI}\gamma_{fE}}{{\cal J}\sqrt{4+\Delta^2}} \,, \\
    {\cal A}_\Sigma(\Omega) \approx {\cal A}_\Sigma \equiv {\cal A} +  {\cal A}_f(0) \label{KLMTVs_A_f_0(b)}\,.
  \end{gather} 
\end{subequations}
It is easy to show, using definitions (\ref{my_crit}, \ref{Omega_c}), that in this case 
\begin{subequations}\label{KLMTVs_xi_Omega}
  \begin{gather}
    \xi^2 = \frac{1}{2}\left(\sqrt{2\Delta^2 + 4{\cal A}_\Sigma} - \Delta\right) \label{KLMTVs_xi2} \,,\\
    \Omega_c^2 = 2{\cal J}\xi^2 \,, \label{KLMTVs_Omega_c} 
  \end{gather}
\end{subequations}
It follows from Eq.\,(\ref{KLMTVs_xi2}) that the optimal value of $\Delta$ has to be small, $\Delta^2\sim{\cal A}_\Sigma\ll 1$. This allows further simplifications:
\begin{subequations}\label{KLMTVs_parms}
  \begin{gather}
    \gamma_{fI} \approx \delta_f \approx \sqrt{{\cal J}} \,, \label{KLMTVs_parms(b)} \\
    {\cal A}_\Sigma \approx {\cal A} + \frac{2\gamma_{fE}}{\sqrt{{\cal J}}} \,, \label{KLMTVs_parms(c)}
  \end{gather}
\end{subequations}
Using Eq.\,(\ref{KLMTVs_Omega_c}), the last formula can be presented as the following:
\begin{equation}
  {\cal A}_\Sigma \approx \frac{1+\Delta^2}{2\xi^2}\,q + 2\xi q_f\,, \label{KLMTVs_A_Sigma}
\end{equation} 
where
\begin{subequations}
  \begin{gather}
    q = \dfrac{mc^2\Omega_c^2A^2}{16\omega_pW} = \frac{2{\cal A}\xi^2}{(1+{\cal A})(1+\Delta^2)}\,, 
      \label{KLMTVs_g} \\
    q_f = \dfrac{cA_f^2}{2\sqrt{2}l_f\Omega_c} \,. \label{KLMTVs_gf} 
  \end{gather}
\end{subequations}
Solution of Eq.\,(\ref{KLMTVs_xi2}) for $q$, with account of Eq.\,(\ref{KLMTVs_parms(c)}), give the following expression:
\begin{equation}
  q = \frac{2\xi^2\left(\xi^4 + \xi^2\Delta -\Delta^2/4 - 2q_f\xi\right)}{1+\Delta^2} \,.
\end{equation} 

If $\Delta=0$, then 
\begin{equation}\label{KLMTV_g_xi_res} q = 2\xi^3\left(\xi^3 -2q_f\right) \,. \end{equation} 
[see also Eq.\,(\ref{KLMTVs_xi_res}).]

The optimal value of $\Delta$ which maximize $q$ (\emph{i.e.} minimize $W$) for given $\xi$, is equal to:
\begin{equation}\label{KLMTVs_optD} 
  \Delta = \frac{1}{\xi^2}\left[
      \sqrt{\left(\xi^4 - 2q_f\xi + 1/4\right)^2 + \xi^4} - \left(\xi^4 - 2q_f\xi + 1/4\right)
    \right]
\end{equation} 
[this detuning corresponds to $\Omega_0\approx\Omega_c$, see Eqs.\,(\ref{FPLs_alpha_beta_X(a)}, \ref{KLMTVs_Omega_c})]. In this case,
\begin{equation}\label{KLMTV_g_xi_det}
  q = \frac{\xi^2}{2}\,
    \frac{3\xi^4 - 2q_f\xi + 1/4 - \sqrt{\left(\xi^4 - 2q_f\xi + 1/4\right)^2 + \xi^4}}
      {\sqrt{\left(\xi^4 - 2q_f\xi + 1/4\right)^2 + \xi^4} - \left(\xi^4 - 2q_f\xi + 1/4\right)}\,.
\end{equation}  
If $\xi\ll 1$, then Eqs.\,(\ref{KLMTVs_optD}, \ref{KLMTV_g_xi_det}) can be simplified:
\begin{gather}
  \Delta\approx2\xi^2 \,, \\
  q \approx \frac{1}{4\xi^3\left(\xi^3 - q_f\right)} \,.
\end{gather}
[see also Eq.\,(\ref{KLMTVs_xi_det})]

\subsection{Estimates}\label{app:estimates}

\begin{table}[t]
  \begin{tabular}{|c|c|c|}
    \hline
            & SFC                                   & LFC \\
    \hline
      $l_f$         & $10\,\mathrm{m}$                  & $4000\,\mathrm{m}$ \\
      $\xi$         & $0.4$                             & $0.1$ \\
      $\Omega_c$    & $2\pi\times500\,\mathrm{s}^{-1}$  & $2\pi\times50\,\mathrm{s}^{-1}$ \\
    \hline
      ${\cal J}$       & $3\times10^7\,\mathrm{s}^{-2}$    & $5\times10^6\,\mathrm{s}^{-2}$ \\
      $q$           & $7\times10^{-3}$                  & $7\times10^{-7}$ \\
      $q_f$         & $0.3$                             & $8\times10^{-4}$ \\
      $\Delta$      & $0.3$                             & $0.02$ \\
    \hline
      $W/m$         & $50\,\mathrm{W/g}$                & $5\,\mathrm{kW/g}$ \\
      ${\cal A}$       & $0.025$                           & $3\times10^{-5}$ \\
    \hline
      $\gamma_{fI}$ & $6\times10^3\,\mathrm{s}^{-1}$    & $2\times10^3\,\mathrm{s}^{-1}$ \\
      $\delta_f$    & $6\times10^3\,\mathrm{s}^{-1}$    & $2\times10^3\,\mathrm{s}^{-1}$ \\
    \hline
  \end{tabular}
  \caption{Two examples of the parameters sets (SFC: short filter cavity; LFC: long filter cavity)}\label{tab:estimates}
\end{table} 

Consider two example sets of the parameters values, which correspond to short ($l_f=10\,\mathrm{m}$) and long ($l_f=4\,\mathrm{km}$) filter cavities. The ``seed'' values are $l_f,\,\xi$ and $\Omega_c$, see Table\,\ref{tab:estimates}. Eqs.\,(\ref{KLMTVs_Omega_c}, \ref{KLMTVs_gf}, \ref{KLMTVs_optD}, \ref{KLMTV_g_xi_det}) give the values of ${\cal J},\,q_f,\,\Delta$, and $q$. Then, Eq.\,(\ref{KLMTVs_g}) allows to calculate $W/m$ and ${\cal A}$. Finally, Eqs.\,(\ref{KLMTVs_parms0}) give the filter cavity parameters $\gamma_{fI},\,\delta_f$ and ${\cal A}_f(\Omega)$ (for $\gamma_{fE}$, see Table\,\ref{tab:notations2}). Substitution of these parameters into Eq.\,(\ref{KLMTV_S}) allows to calculate the sum noise spectral density. The result is presented in Fig.\,\ref{fig:varmeas2}. 

It have to be noted that in Eq.\,(\ref{KLMTV_S}), detuning $\Delta$ and spectral dependence of ${\cal A}_f(\Omega)$ are taken into account. Nevertheless, the resulting spectral densities correspond to the ``seed'' values $\xi=0.4$ and $\xi=0.1$ with good precision, which verify approximations made above.


\end{document}